\documentclass[10pt,journal,compsoc]{IEEEtran}
\ifCLASSOPTIONcompsoc
  \usepackage[nocompress]{cite}
\else
  \usepackage{cite}
\fi
\ifCLASSINFOpdf
\else
\fi
\usepackage[utf8]{inputenc}
\usepackage{titlesec}
\usepackage{graphicx}
\usepackage{url}
\usepackage{courier}
\usepackage{multicol}
\usepackage{multirow}
\usepackage{comment}
\usepackage[justification=centering]{caption}
\usepackage{footnote}
\usepackage{xcolor}
\usepackage{pifont}
\newcommand{\cmark}{\text{\ding{51}}}
\newcommand{\xmark}{\text{\ding{55}}}
\usepackage[many]{tcolorbox}    
\newtcolorbox{mybox}[1]{%
    tikznode boxed title,
    enhanced,
    arc=0mm,
    interior style={white},
    attach boxed title to top center= {yshift=-\tcboxedtitleheight/2},
    fonttitle=\bfseries,
    colbacktitle=white,coltitle=black,
    boxed title style={size=normal,colframe=white,boxrule=0pt},
    title={#1}}

\begin{document}
\title{Let's Go to the Whiteboard (Again):\\
       Perceptions from Software Architects\\
       on Whiteboard Architecture Meetings}

\author{Eduardo Santana de Almeida,~\IEEEmembership{Senior Member,~IEEE,}
        Iftekhar Ahmed,
        and~Andr\'e van der Hoek,~\IEEEmembership{Senior Member,~IEEE}% <-this % stops a space
\IEEEcompsocitemizethanks{\IEEEcompsocthanksitem E. Almeida was with the Institute of Computing (IC-UFBA), Federal University of Bahia, Brazil.\protect\\
E-mail: esa@rise.com.br
\IEEEcompsocthanksitem I. Ahmed and A. van der Hoek are with University of California, Irvine (UCI).}% <-this % stops an unwanted space
\thanks{Manuscript received ...; revised...}}


\markboth{Journal of \LaTeX\ Class Files,~Vol.~14, No.~8, August~2015}%
{Shell \MakeLowercase{\textit{et al.}}: Bare Demo of IEEEtran.cls for Computer Society Journals}



\IEEEtitleabstractindextext{%
\begin{abstract}
  %"Hey, Folks. I designed the architecture of our solution. Let's go to the whiteboard and I can share with you the main decisions taken and clarify some doubts". 
  %It is a typical day in a week of a software architect. 
  The whiteboard plays a crucial role in the day-to-day lives of software architects, as they frequently will
  organize meetings at the whiteboard to discuss a new architecture, some proposed changes to the architecture, a
  mismatch between the architecture and the code, and more. 
  %in their day-to-day work. go to the whiteboard whiteboard software architecture meetings to design, discuss, and evaluate the software architecture with members of the development team. 
  While much has been studied about software architects, the architectures they produce, and how they produce them,
  a detailed understanding of these whiteboards meetings is still lacking. In this paper, we contribute a mixed-methods study involving semi-structured interviews and a subsequent survey to understand the perceptions of software architects on whiteboard architecture meetings. We focus on five aspects: (1) why do they hold these meetings, what is the impact of the experience levels of the participants in these meetings, how do the architects document the meetings, what kinds of changes are made after the meetings have concluded and their results are moved to implementation, and what role do digital whiteboards plays?  In studying these aspects, we identify 12 observations related to both technical aspects and social aspects of the meetings. These insights have implications for further research, offer concrete advice to practitioners, provide guidance for future tool design, and suggest ways of educating future software architects.  
\end{abstract}

}

\maketitle

\IEEEdisplaynontitleabstractindextext

% \fi
\IEEEpeerreviewmaketitle

\section{Introduction}
\label{sec:intro}

%{\textcolor{red}{ I removed the first paragraph.}}
%In the 1990s, software architecture emerged as an important subdiscipline of software engineering, especially in the context of large-scale systems \cite{Perry:1992:FSS:141874.141884}. The principled use of software architecture can have a positive impact on various aspects of software development, including ease of understanding, reuse of components, evolution and maintenance, analysis of system properties, and overall management of the development project~\cite{?}. At the same time, designing or choosing an inappropriate architecture can have serious negative consequences \cite{Garlan:1995:ISI:205313.205314}.

Designing a software architecture is not purely a technical issue \cite{Fowler:2003:NA:942589.942716, DBLP:conf/wicsa/ClementsKKDRV07}. It also involves numerous social, human, and organizational aspects \cite {paper:conway:1968} that can influence the success of an architecture, and thus the entire project, considerably. One such aspect concerns who is involved in the design process: the different stakeholders participating must be selected with care~\cite{Brooks:1987:NSB:26440.26441, DBLP:journals/software/PerrySV94}. So it is for the primary software architect or architects leading the effort: they are typically experienced members of the team who are ultimately responsible for the design choices made, validating them, and capturing and sharing them in various kinds of artifacts that will be used downstream~\cite{KRUCHTEN20082413}. 

To date, the software engineering community has studied which kinds of projects and organizations need a software architect \cite{Fowler:2003:NA:942589.942716}, the assigned duties, skills, and knowledge of architects \cite{DBLP:conf/wicsa/ClementsKKDRV07}, what software architects actually do \cite{KRUCHTEN20082413, DBLP:journals/jss/HoornFLV11}, their mindset \cite{DBLP:conf/qosa/ClercLV07}, the reasoning process involved in making architectural decisions \cite{DBLP:conf/wicsa/HeeschA11, DBLP:conf/ecsa/PowerW18}, and the impact of software architects writing code themselves \cite{DBLP:journals/software/Woods17, DBLP:conf/icse/RehmanMNUT18}, among others. Not all findings from these studies are uniform. For instance, what architects do varies considerably from one organization to another and even from one project to another within the same organization \cite{Taylor:2009:SAF:1538494, DBLP:conf/icse/RehmanMNUT18}. Moreover, gaps exist in our collective understanding, which has led to calls for further studies of software architects, their work, and how they conduct it \cite{DBLP:journals/ese/FalessiBCK10, DBLP:journals/software/AntoninoMK16, DBLP:journals/software/ErderP16}.

%since we do not have a sufficient understanding, for example, of what novice and experienced software architects do when they design, evaluate, and document a software architecture. 

This paper contributes one such study, with two closely related  goals. Our first goal is to understand how software architects perceive the whiteboard design meetings in which they participate to engage in high level architectural design. Whiteboards are an important medium used in reasoning and problem solving in many disciplines, such as civil engineering \cite{Do1996ReasoningAC}, mechanical engineering \cite{Henderson:1998:LPV:521215}, and design engineering \cite{Henderson:1998:LPV:521215}. In software engineering, previous whiteboard studies focused on programmers \cite{DBLP:journals/software/PerrySV94, LaToza:2006:MMM:1134285.1134355, Cherubini:2007:LGW:1240624.1240714} and software designers \cite{Mangano:2014:SID:2556288.2557411, DBLP:journals/tse/ManganoLPH15}, but not architects. 
%While architecture generally is considered design, not all design is considered architecture.\footnote{https:\/\/handbookofsoftwarearchitecture.com} 
With software architecture design laying the critical foundations for a project's success, then, it is important to specifically study software architecture design work being performed at the whiteboard to understand the unique context, demands, and approaches of this setting. 

Our second goal is to understand the transition from architectural design sketches as produced in whiteboard software architecture meetings to implementation. Architecture work at the whiteboard is necessarily incomplete and, when the rubber hits the road during more detailed design or implementation, refinements and changes are frequently needed~\cite{DBLP:journals/tse/ManganoLPH15}. While much has been said about the need for architecture and implementation to stay in sync~\cite{DBLP:journals/ese/SharmaSS20}, as well as about architectural erosion and drift over time~\cite{DBLP:journals/smr/LiLSA22}, no study to date has examined how information from whiteboard architecture design meetings makes its way into implementation, what kinds of deviations happen in this process, and why.

%Sketches depict parts of the mental model developers build to understand a software project \cite{10.1145/2635868.2635891}. They can contain different views, levels of abstraction, formal and informal notations and pictures \cite{Cherubini:2007:LGW:1240624.1240714, Mangano:2014:SID:2556288.2557411}. In whiteboard architecture meetings, different kinds of design decisions are taken. Thus, is important to know how software architects document these decisions, which types of sketches are created, and the changes that happen from sketches at whiteboard to implementation. 

%Over a period of ten months, 
We conducted semi-structured interviews with 27 software architects from 18 different companies across five different countries, with the software being worked on by these architects spanning a wide range of application domains. We then surveyed an additional 46 software architects across nine countries to contextualize and augment the findings from the interviews. 

The following five research questions guided the analysis of the interviews and survey responses were 
\begin{itemize}
    \item \textbf{RQ1:} Why do software architects hold whiteboard software architecture design meetings?
    \item \textbf{RQ2:} What is the impact of the experience of the participants on how design work 
              proceeds in these meetings?
    \item \textbf{RQ3:} How do software architects document what happens in whiteboard 
              software architecture meetings?
    \item \textbf{RQ4:} When moving to implementation, what kinds of refinements and
              deviations happen to the architectural design that was first
              created?
    \item \textbf{RQ5:} What role do digital whiteboard tools play?
\end{itemize}
This paper constitutes the first broad empirical study of how software architects view and engage in whiteboard architecture meetings, as well as the outflow of those meetings into eventual code.  It makes the following contributions:
\begin{itemize}
\item A mixed qualitative and quantitative study that investigates key aspects of whiteboard software architecture meetings, with a primary focus on the role of experience and the transition of the results from these meetings to implementation. 
%Our present results from analyzing 27 interviews with software architects as well as the survey responses of 46 additional software architects.
\item A set of twelve observations regarding the nature of whiteboard software architecture meetings, covering  concrete technical, human, and social perspectives on these meetings and offering recommendations for practitioners, researchers, tool builders, and educators. 
\item A collection of all our research materials on a project website
%\footnote{https://github.com/whiteboard-architecture/empirical-study} 
for replication and reproducible research purposes, including our interview data (prompts, transcriptions, and codebook) and the survey instrument.
\end{itemize}

\section{Related Work}

\subsection{Studies of Sketching and Whiteboard Use}

Dekel and Herbsleb conducted an observational study analyzing several software design meetings from the ACM DesignFest event that was held at the 2005 OOPSLA conference~\cite{DBLP:conf/oopsla/DekelH07}. The analysis of the meetings primarily focused on the notations that the designers used and the representations they created in those notations, detailing for instance how the designers often started with unstructured representations and how they at times combined content from what were independently created diagrams into a single diagram. Based on the results of the study, Dekel and Herbsleb discussed several implications for the design of future tools supporting whiteboard based design. Cherubini et al. \cite{Cherubini:2007:LGW:1240624.1240714} performed an exploratory study at Microsoft of how and why developers use whiteboards, with a particular focus on how developers `draw their code'. Based on interviews and surveys with developers, they found that informal notations were used in support of face-to-face communication about the code and that available modeling tools were not capable of supporting this need since their focus on formal, correct diagrams does not match the informal nature in which developers seek to externalize their mental models of code. 
%They found also that the role of sketches in software development differs from other engineering disciplines. 

Leveraging one of the videos collected for the 2009 Studying Professional Software Design workshop \cite{10.5555/2535028}, Nakakoji et al. examined the conversations and whiteboard drawings of a pair of professionals with the help of the Design Practice Streams (DPS) tool they developed~\cite{6035659}. DPS allows replays of strokes on the whiteboard and connects the replay to an automatically created transcript of the meeting, enabling quick exploration of concepts and when they were talked about. In the case of the video analyzed, Nakakoji et al. highlight the role of key concepts and what aspects of the design being worked on were most frequently re-discussed.

In order to understand how to provide improved tool support for integrating visual sketches into developers' workflows, Walny et al. conducted a qualitative study centered on the creation, use, and transformation of sketches~\cite{6069462}. Using semi-structured interviews with eight software developers, their particular focus was on the lifetime of sketches: with what medium they were created first (e.g., paper, whiteboard, tool) and how they then were captured, augmented, and re-created in similar and other media (e.g., photo, tablet, another piece of paper).

Baltes and Diehl investigated the use of sketches and diagrams in software engineering practice, with a particular focus on their relation to source code artifacts~\cite{10.1145/2635868.2635891}. Using data from three companies and a survey, they found that the majority of the sketches were informal and that the most common purposes for creating sketches and diagrams were designing, explaining, and understanding. More than half of the sketches were created on analog media like paper or whiteboards and were revised after creation. Based on the findings from this work, Baltes et al. developed SketchLink, a tool that aims at increasing the value of sketches and diagrams by explicitly linking them to the source code to which they pertain~\cite{10.1145/2635868.2661672}. 

Mangano et al. conducted an observational study analyzing fourteen hours of design activity by eight pairs of professional software developers at the whiteboard~\cite{DBLP:journals/tse/ManganoLPH15}. In the study, each pair was provided with a written prompt asking them to design an educational traffic light simulation program. The researchers analyzed the type of sketches created, how professional software designers focus on individual sketches and shift their attention among sketches, and the reasoning process to understand and advance the state of the design at hand. 

Out of these prior studies, the ideas discussed in \cite{Cherubini:2007:LGW:1240624.1240714}, \cite{Mangano:2014:SID:2556288.2557411}, and \cite{10.1145/2635868.2635891} are most closely related to our study. However, our study is unique in focusing on software architects and software architecture design activities at the whiteboard, as well as in examining important aspects not covered in previous work, such as the influence of levels of experience and the transition from sketches at the whiteboard to code. 

Beyond studies that specifically focus on creating an understanding of sketching and whiteboard use, many tools have been proposed to explicitly support software developers in their design sketching. Baltes et al. \cite{8091190}, for instance, presented LivelySketches, a prototype tool that supports the round-trip lifecycle of sketches from analog to digital and back. As another example, FlexiSketch supports collaboration across distances, with the ability for users to define notations on the fly~\cite{DBLP:conf/icse/WuestSG15}. An interesting recent example seeks to integrate sketching in the IDE~\cite{DBLP:conf/icse/SamuelssonB20}. Compared to these and many other tools that have been proposed, our paper does not contribute any tool designs, though our findings give rise to implications for future tools.

\subsection{Studies of Software Architects}

\textbf{Personal Experiences.} The Pragmatic Architect column in the IEEE Software magazine discussed many aspects about the software architect and their role over the years. In \cite{Fowler:2003:NA:942589.942716}, for instance, Fowler introduced key definitions of software architecture and the architect's role in creating and managing them. Buschmann \cite{DBLP:journals/software/Buschmann12a} shared and commented on a compressed week-long diary of what his `real life' as an architect is like when working on a product line for an industrial automation system. Based on more than ten years working as a software architect, Woods \cite{6802995} classified architects into three groups: enterprise, infrastructure, and application architects. Klein \cite{DBLP:journals/software/Klein16} discussed what makes a software architect successful. Erder and Pureur \cite{DBLP:journals/software/ErderP16} considered the architect's role in agile development and followed up on this discussion in a later paper that examined desired personality traits of software architects~\cite{DBLP:journals/software/ErderP17}. Woods \cite{DBLP:journals/software/Woods17} discussed the benefits and drawbacks of architects actually engaging in programming beyond their primary role as a designer. Klein \cite{DBLP:conf/wicsa/Klein05}, based on his experience in industry, defined a three-phase model (Blank Page, Integration, and Magic) to capture the evolution of software systems, and discussed the kinds of contributions necessary from the software architect for achieving success in each phase.  

Sarang \cite{DBLP:conf/wicsa/Sarang07} proposed a structure for an architecture team and defined the roles and responsibilities of the members. Based on his experience in managing a 10-person architecture team from 1992-1995, Kruchten \cite{KRUCHTEN20082413} described what software architects `really do'.

\textbf{Surveys.} Clements et al. \cite{DBLP:conf/wicsa/ClementsKKDRV07} investigated the human aspects of architecting software, focusing on the duties, skills, and knowledge of software architects. They canvased over 200 public sources of information (e.g., web sites, blogs, training and education materials, job descriptions) to identify about 200 different duties, 100 skills, and 100 areas of knowledge -- each of which was mentioned by at least one source.

Clerc et al. \cite{DBLP:conf/qosa/ClercLV07} performed a survey in the Netherlands to collect feedback on the importance of architectural knowledge for the daily work of practitioners in architecture. Based on the answers of 107 respondents, the study provides insights in the way practitioners view and use architectural knowledge by listing what uses are important for the different roles that architects play (e.g., project lead, reviewer, consultant) and on what architectural level (e.g., software, information, enterprise). 

Heesch and Avgeriou \cite{DBLP:conf/wicsa/HeeschA11} surveyed 53 software architects from several companies and project domains to get insights in the reasoning processes followed in architectural design. Among a variety of findings, they show that architects typically are involved in requirements elicitation and therefore understand the reasoning behind the requirements well, that architects find it important to search for multiple options but equally consider this an expensive activity and only engage in doing so when truly necessary (thus favoring known solutions), and that architects seldom reject decisions they have made before. 

Hoorn et al. \cite{DBLP:journals/jss/HoornFLV11} conducted a large-scale survey with 142 software architects from four IT organizations in the Netherlands to understand what architects do on a day-to-day basis and what kind of support they need for sharing architectural knowledge. 

\textbf{Case Studies.} Premaj et al. \cite{DBLP:conf/wicsa/PremrajNTV11} conducted a case study with two projects by performing retrospective root cause analyses into the issues assigned to software architects to understand why the issues arise, what types they are, and how their occurrences could potentially be reduced in future through improvements in the development process.  

Rehman et al. \cite{DBLP:conf/icse/RehmanMNUT18} conducted a two-stage case study, combining data analytics for five open source projects with semi-structured interviews of several architects of these systems. The authors addressed three questions: Do architects write code? What type of code do architects write? Is there any empirical evidence to support that software projects will benefit from hands-on software architects? the primary conclusion is that benefits exist to architects writing code.
 
While the focus of our study is different from these existing studies of software architects, the viewpoints expressed in the personal experiences and the results from the case studies were particularly influential in shaping the direction of our study to provide a complementary view of an important under-understood activity: whiteboard architectural design together with its downstream outflow. In addition, the surveys conducted in \cite{DBLP:conf/wicsa/ClementsKKDRV07}, \cite{DBLP:journals/jss/HoornFLV11}, \cite{DBLP:conf/qosa/ClercLV07}, \cite{DBLP:conf/wicsa/HeeschA11},  \cite{DBLP:conf/wicsa/PremrajNTV11}, and \cite{DBLP:journals/jss/HoornFLV11} served as inspiration for how we structured our survey questions.

\section{Research Design}

We adopted a two-part research design for our study: we first conducted in-depth semi-structured interviews with experienced software architects and then performed a validation survey with additional, again experienced, software architects. The goal of the interviews was to gather insights into practitioner views to help us to formulate a clear picture of whiteboard software architecture meetings and the outflow from these meetings. These insights were then checked and refined by the results from the validation survey. All study materials from the interviews and surveys can be found in the supplementary materials for the paper.\footnote{https://github.com/whiteboard-architecture/empirical-study}

\subsection{Interviews}

\textbf{Protocol.} We interviewed 27 software architects with experience in whiteboard software architecture meetings. The first author interviewed the software architects either in person, if they worked in the same area, or via Skype, if they did not. The average length of the interviews was 37 minutes, with the shortest interview 7.18 minutes and the longest 43.42 minutes.  

Each interview consisted of two parts. In the first part, in addition to a few demographic questions and questions about the experience level of the interviewee, the interviewer asked open-ended questions about their engagement in whiteboard architecture design meetings, their perceptions about the participation of novice and experienced software architects, and their advice for other software architects participating in these kinds of meetings. 
In the second part, the interviewer asked questions related to the importance of the meetings to implementation, approaches used to document and communicate outcomes, decisions and structures preserved from whiteboard to code, typical changes that take place when moving from sketched designs to concrete implementation, and aspects missing from whiteboard discussions. Finally, we thanked the interviewees and debriefed them by informing them about what we planned to do with the data. The protocol was designed by two researchers over a period of three months.

\textbf{Participants.} We selected the software architects to be interviewed based on convenience sampling and snowballing. Initially, we invited twenty software architects with experience in designing software architectures and participating in whiteboard architecture meetings. All of them agreed to be interviewed and they recommended seven additional software architects to be interviewed as well. These seven recommended software architects also agreed to participate in the study.

We first conducted a pilot interview with another software architect (not included in the study) as a pretest \cite{Seidman2006} and then performed the interviews with the 27 architects. The software architects came from eighteen different companies across five different countries (Brazil 13, Canada 1, Germany 4, Sweden 2, and USA 7), working across a range of different domains, including but not limited to games, music, finance, e-commerce, data science, and sports. 
%Qualitative researchers \cite{Strauss2007} advise that at least ten interviews or observations with detailed coding are necessary for building a grounded theory (p. 281). Our goal was not to build a ground theory and 
Note that after 24 interviews, we reached saturation. We performed a few extra interviews to ensure this was the case, reaching a total of 27.
%after the 27 interviews, we reached saturation. 
All interviews were conducted by the first author over a period of four months. Only one interview was completed face to face.  
 
In the remainder of the paper, we label each interviewee SA1 through SA27. Table \ref{table:tablebackground} provides a summary of the architects' backgrounds. The supplementary material attached to the paper presents detailed (though anonymized) information about the participants.

\begin{table}[h!]
\centering
\begin{tabular}{||c c c c||} 
 \hline
\textbf{SA} & \textbf{Experience} & \textbf{Company sector} & \textbf{Country}  \\ 
 \hline\hline
 SA1 & 2-5 years & IT & Brazil \\ 
 SA2 & 5+ years & e-Government & Brazil \\ 
 SA3 & 2-5 years & e-Commerce & Brazil \\ 
 SA4 & 5+ years & IT & Brazil \\
 SA5 & 5+ years & IT & Brazil \\
 SA6 & 2-5 years & Marketing & Brazil \\ 
 SA7 & 5+ years & Audio streaming & Sweden \\
 SA8 & 5+ years & IT & Brazil \\
 SA9 & 5+ years & IT & Brazil \\
 SA10 & 5+ years & Finance & Brazil \\
 SA11 & 5+ years & Marketing & Brazil \\
 SA12 & 2-5 years & Embedded Software & Germany \\
 SA13 & 5+ years & Finance & Brazil \\
 SA14 & 5+ years & e-Commerce & Canada \\
 SA15 & 5+ years & Embedded Software & USA \\
 SA16 & 5+ years & e-Government & Brazil \\
 SA17 & 5+ years & Embedded Software & USA \\
 SA18 & 5+ years & e-Government & Brazil \\
 SA19 & 5+ years & Games & Sweden \\
 SA20 & 5+ years & Cyber-physical & USA \\
 SA21 & 5+ years & Embedded Software & Germany \\
 SA22 & 5+ years & Embedded Software & USA \\
 SA23 & 5+ years & Embedded Software & Germany \\
 SA24 & 5+ years & Embedded Software & USA \\
 SA25 & 5+ years & Embedded Software & Germany \\
 SA26 & 5+ years & Embedded Software & USA \\
 SA27 & 5+ years & Sports & USA \\
 \hline
\end{tabular}
\caption{Summary of Professional and Demographic Information of Participants.}
\label{table:tablebackground}
\end{table}

\textbf{Data analysis.} Data analysis started with audio transcription (16 hours and 46 minutes total). Two researchers conducted the transcription using the Trint tool\footnote{https://trint.com} and, after internal review, all interview transcripts were formatted and then shared with the respective software architects for validation; this resulted in minor adjustments involving the architects providing clarifications of some of their answers. 

After that, the coding process was started by two researchers using the NVivo tool\footnote{https://www.qsrinternational.com/nvivo/nvivo-products}. The two researchers had previous experience in conducting studies combining interviews and surveys, albeit in different aspects of software engineering. We used a set of first cycle and second cycle coding methods to data analysis \cite{saldana2015coding}. First cycle methods are those processes that happen during the initial coding of data. Second cycle methods, if needed, are advanced ways of reorganizing and reanalyzing data coded by first cycle methods. In our study, the codes created in the first cycle were clustered in categories in the second. These clusters, then, served as the source of our results as discussed in Section~\ref{sec:results}).
%Based on our data, we considered that the categories defined were appropriated to understand the answers that the interviews participants gave. 
Note that we do not strive to develop a theory at this time. Doing so is not always a necessary outcome for a qualitative study \cite{mason} and, in our case, is actually ill suited for the research questions we pose. Rather, then, we seek to document the perceptions of software architects in a full and diverse light.
 
Once each researcher had performed their independent turn, the two authors met to resolve differences in coding. %After discussing and comparing, we established a ``substantial agreement'' of 0.65 measured using Cohen's kappa~\cite{landis1977measurement}.

%a high inter-rater agreement of 70\%, which according to Landis et al.~\cite{landis1977measurement}, is considered as substantial level of agreement. 

\subsection{Survey}

\textbf{Protocol.} Based on the results from the interviews, we designed a 30-minute survey to further build our understanding of the perceptions of software architects on whiteboard software architecture meetings. The survey was composed of seventeen questions, fifteen of which were closed questions and two of which were open questions. The survey also collected demographic information from respondents. For the design of the survey, we followed Kitchenham and Pfleeger's guidelines for personal opinion surveys \cite{BarbaraKitchenham_2008}. As one of the guidelines, previous surveys related to sketches in software engineering \cite{10.1145/2635868.2635891} and software architects \cite{DBLP:conf/wicsa/ClementsKKDRV07, DBLP:conf/qosa/ClercLV07, DBLP:conf/wicsa/HeeschA11, DBLP:journals/jss/HoornFLV11} were consulted.

We piloted our survey with two experienced software architects to get feedback on the formulation of the questions, difficulties faced in answering the survey overall, and time to finish it. As these pilot respondents were experts in the area, we also wanted to know whether they felt we were asking the right kinds of questions or should be changing the approach. In response to their feedback, we modified the survey several times, rephrasing some questions and removing others to make it easier to understand and answer. The final version of the survey consisted of 23 questions (including demographics). 
%Another concern in this stage was also to ensure that the participants could finish it in 30-minutes. 
The pilot survey responses were used solely to improve the questions, and these responses were not included in the final results. We kept the survey anonymous, but in the end, the respondents could share their email to receive a summary of the study. The survey instrument is included in the supplementary material.

\textbf{Participants}. We followed a three-step approach to recruit survey respondents: initially, we posted survey information on personal accounts on social media (e.g., Twitter, LinkedIn). Next, two authors contacted potential respondents by email (convenience sample) and asked them to share it with other potential respondents (snowballing). Because of this process, we were not able to track the total number of invitations. Overall, we received 50 responses, out of which we disqualified four responses that did not have any responses to any of the actual survey questions of interest (despite having responses to the basic demographics questions). This lead to 46 valid responses that were considered where the survey respondents answered all questions.

The respondents were spread out across nine countries and  four continents. The top three countries where the respondents came from were United States, Brazil, and Germany. The professional experience of the 46 respondents working as software architects varied from one year to 31 years, with an average of 14 years and a median of 15 years. The majority of the respondents had an advanced degree (67.4\%), i.e., Master's or Ph.D., 30.4\% of the respondents had a Bachelor's degree, and 2.2\% graduated from high school without completing college.

\textbf{Data Analysis}. We collected the ratings that our respondents provided for each closed question and converted these ratings to Likert scores from 1 (Strongly Disagree) to 5 (Strongly Agree). We computed the average Likert score of each statement related to different perspectives (e.g., reasons to conduct whiteboard architecture meetings, different levels of experience, role of documentation, transition from sketches to code, and digital tools) and plotted Likert scale graphs. In addition, we used open coding to analyze the answers that the survey respondents gave to the two open questions related to recommendations for  software architects conducting whiteboard architecture meetings and final thoughts on the topic. To reduce subjective bias during the open coding process, we assigned both to two authors of this paper. Each author analyzed the answers separately. Once all the data were coded, the two authors met to resolve differences in coding.

\section{Results}
\label{sec:results}

In this section, we present the results for each of the five research questions identified in Section~\ref{sec:intro}. Before we do so, however, we first further characterize whiteboard software architecture meetings based on the responses from the architects.

\subsection{Whiteboard Software Architecture Meetings}

%Before starting discussing different perspectives on whiteboard software architecture meetings, it is important to characterize them. 
During the interview sessions, we asked the participants to give an example or two of recent whiteboard software architecture meetings in which they participated. We specifically asked them to describe the setting, how many participants there were, what roles they held, what problem the group addressed, and how long the meeting lasted.

The meetings involved from two (min) to 15 (max) participants, with five being the average. Beyond the interviewed software architects themselves, other participants held many different roles, including other software architects, developers, requirements engineers, test engineers, marketing managers, product managers, consultants, UX engineers, and users. According to one of the software architects (SA 10): \textit{"... you have a team and everyone is working together, so the developer who is doing the code is participating in the architecture discussion. The tester is also participating, the requirements analyst too, everyone is on the same boat. There is no longer separation."}

The topics discussed ranged from system integration and service design, through cloud deployment and performance, to knowledge management and even pre-sales. As one example, a software architect (SA 21) described a whiteboard meeting concerning the performance of the software for which they are responsible as follows: \textit{"We're using Lambda functions to go from one step to another and we are seeing performance challenges on processing large files given the limitation of the Lambda function of having five hundred MB per data frame, so you can open any file that will consume less than half GB memory of RAM and then do the processing."} As another example, another architect (SA 7) described a meeting they organized as: \textit{"As we had a system that exchanged many messages over the network and needed very high performance, several times, we discussed architectural problems that could bring I/O bottlenecks, because it was a system that wrote a lot on disk and sent many messages over the network, so we had I/O bottlenecks on the machine running, as network bottlenecks needed high-level architectural discussions"}.

The meetings ranged from 20 minutes (min) to seven hours (max), with 1 hour and 10 minutes being the average duration. According to one of the software architects (SA 7), meetings take longer when they involve activities that expressly seek to document outcomes: \textit{"These meetings are roughly I would say between four and seven hours. And this other meeting that I told you about where we are documenting. So, when do we do that, we typically also reserve at least half a day or even better a day. So that we spend also at least four to six hours really working on it."} This clearly is toward the extreme end of length of meeting; at the same time, it recognizes that architecture work is not easy and requires participants to engage in depth to work through what sometimes can be very complex issues. In this case, the culture at the company at which the architect works is such that longer meetings to sort things through are preferred to spreading out the work over multiple, more disconnected meetings.

\subsection{Reasons to Conduct Whiteboard Software Architecture Meetings (RQ1)}

\begin{figure*}[h!]
    \centering
    \includegraphics[width=18cm]{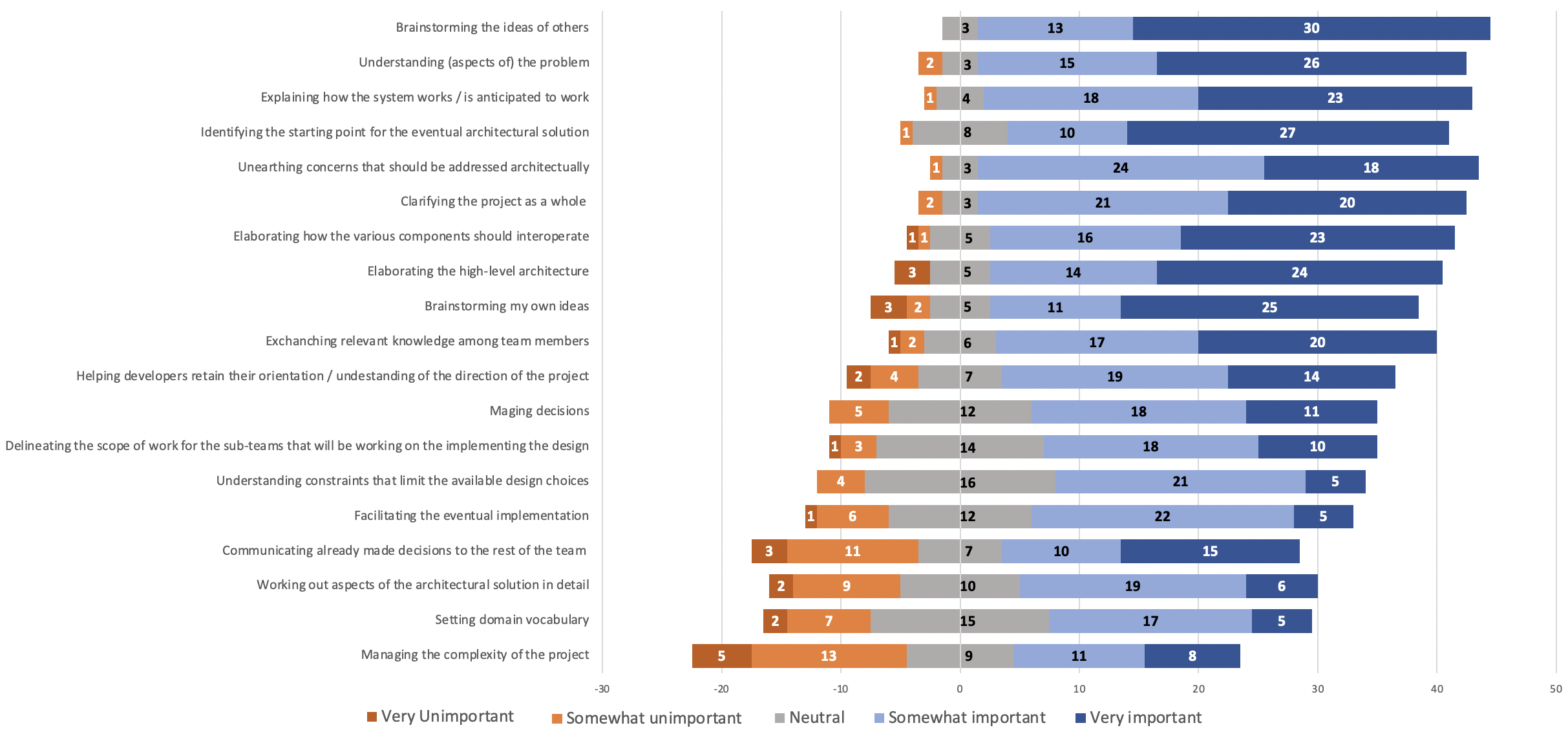}
    \caption{Nineteen Reasons for Software Architects to Conduct Whiteboard Software Architecture Meetings.}
    \label{fig:galaxy}
\end{figure*}

From the interviews, we identified 19 different reasons for software architects to conduct whiteboard architecture meetings. We used the survey responses to rank these 19 reasons. Figure \ref{fig:galaxy} shows the respondents' answers for the nineteen reasons to software architects conduct whiteboard architecture meetings. The top five reasons are: \textit{brainstorming the ideas of others} (average Likert score for this statement is 4.59, i.e., between "somewhat important" and "very important"), \textit{understanding (aspects of) the problem that the architecture has to solve} (average Likert score for this statement is 4.54, i.e., between "somewhat important" and "very important"), \textit{explaining how the system works/is anticipated to work} (average Likert score for this statement is 4.37, i.e., between "somewhat important" and "very important"), \textit{identifying the starting point for the eventual architectural solution} (average Likert score for this statement is 4.37, i.e., between "somewhat important" and "very important"), and \textit{unearthing concerns that should be addressed architecturally} (average Likert score for this statement is 4.28, i.e., between "somewhat important" and "very important"). 

The following are some comments from our software architects (SA) that highlight these aspects, together with what reason it was classified as:

\cmark \textit{SA 3: "Most of the time that an important decision was to be made or some doubt existed related to some important decision, we used the whiteboard to discuss or brainstorm things or draw some things that were the common understanding of the team.}". [understanding (aspects of) the problem that the architecture has to solve].

\cmark \textit{SA 13: "So I would say that the solution crystallizes on the whiteboard. The whiteboard makes it possible to quickly change the solution or to brainstorm some different aspects. So, the brainstorming is I would say impossible without the whiteboard}". [brainstorming the ideas of others].

%In same way, our results were similar to previous work. In \cite{Cherubini:2007:LGW:1240624.1240714}, Cherubini et al. identified nine scenarios where developers employed drawings. The three main reasons why developers produced visualizations were: to understand, to design and to communicate. 
%

In our survey, we also asked respondents about the importance of whiteboard architecture meetings to successful architectural design. Among the 46 survey respondents, 56.52 percent (26 respondents) consider whiteboard design meetings very important to successful architectural design, 41.30 percent (19) ranked the meetings as important, and 2.17 percent (1) were neutral.
%26, 19, and 1 respondents consider very important, important, and neutral this statement.

\begin{mybox}{Observation 1}
Brainstorming the ideas of others, understanding (aspects of) the problem that the architecture has to solve, explaining how the system works/is anticipated to work, identifying the starting point for the eventual architectural solution, and unearthing concerns that should be addressed architecturally are the top five reasons for software architects to conduct whiteboard software architecture meetings.
\end{mybox}

\subsection{Experience (RQ2)}
\label{subsec:experience}

\subsubsection{Important Aspects of Experience}
According to Kruchten \cite{KRUCHTEN20082413}, a software architect is a software expert responsible for designing, developing, nurturing, and maintaining the architecture of the software-intensive systems in which they are involved. Kruchten further observes that, in general, the architect's role is typically reserved for someone with significant experience in prior projects. 

In order to better understand what kind of experience is relevant to whiteboard software architecture meetings, we asked the software architects during our interviews. Collectively, they identified sixteen different aspects, as listed in Table \ref{table:1} with the number of architects that identified each aspect.  The top five aspects are: \textit{design of previous architectures} (5), \textit{ability to communicate} (4), \textit{participation in previous projects} (3), \textit{technical knowledge} (3), and \textit{domain knowledge} (3). 
%The numbers represent the amount of architects who mentioned it. Table \ref{table:1} shows the interviewees' answers for the seventeen aspects related to experience and its frequency.

The following are some comments from some of the software architects that highlight these aspects, together with the corresponding meaning of experience:

\cmark \textit{SA 9: "When we are with experienced architects, they have already designed and implemented various software and have more notion of what works and does not work, they have already seen several examples of architecture, so that is what they bring of importance, the experience itself.}" [design of previous architectures]

\cmark \textit{SA 28: "So I think there is probably two aspects to the experience, right. One is: experience with how to communicate, right. So how to, how to produce design on the fly and communicate that to the other people in the room quickly.}" [ability to communicate]

\cmark \textit{SA 7: "Experience is also defined in terms of the variety of projects. If a person has developed an embedded system, a mobile system, a web system, a large-scale micro-controller system, in various contexts, [ed: they] will be able to compare very well the performance of a web system is different from mobile performance, embedded, etc.}" [participation in previous projects]

\begin{table}[h!]
\centering
\begin{tabular}{||p{6.3cm} c ||} 
 \hline
\textbf{Meaning of Experience} & \textbf{Frequency} \\ [0.5ex] 
 \hline\hline
 Design of previous architectures & 5  \\ 
 Ability to communicate & 4 \\
 Participation in previous projects & 3 \\
 Technical knowledge & 3 \\
 Domain knowledge & 3 \\ 
 Architectural knowledge on methods and tools & 2 \\ 
 Awareness of technology trends & 1 \\ 
 Ability to think strategically & 1 \\ 
 Blend of technical and non-technical aspects & 1 \\ 
 Depth of knowledge about a theme with breadth of knowledge overall & 1 \\ 
 Facilitation of meeting discussion & 1 \\ 
 Having faced failures & 1 \\ 
 Good understanding about existing systems & 1 \\ 
 Ability to interact with project manager & 1 \\ 
 Many hours of development & 1 \\
 Knowing a lot of abstractions across domains & 1 \\
 \hline
\end{tabular}
\caption{Important Aspects of Experience Relevant to Whiteboard Software Architecture Meetings.}
\label{table:1}
\end{table}

Overall, while the aspects mentioned differ, we note that the majority point to the need for a strong technical background that is not necessarily limited to just one project or type of project. 

\begin{mybox}{Observation 2}
Design of previous architectures, ability to communicate, participation in previous projects, technical knowledge, and domain knowledge are the top five aspects of experience seen as useful to whiteboard software architecture meetings.
\end{mybox}

\subsubsection{Team Composition.} Most companies have a limited number of software architects and depending on the size of the company, it may have just one or two people in this role. Thus, whiteboard software architecture meetings necessarily not only involve participants in different roles as discussed prior, but also participants who are likely to have different levels of experience. In the interview sessions, we asked about these different levels of expertise as they relate to team composition for meetings at the whiteboard, and we used the surveys to gather additional information in this regard.

\begin{figure*}[h!]
    \centering
    \includegraphics[width=18cm]{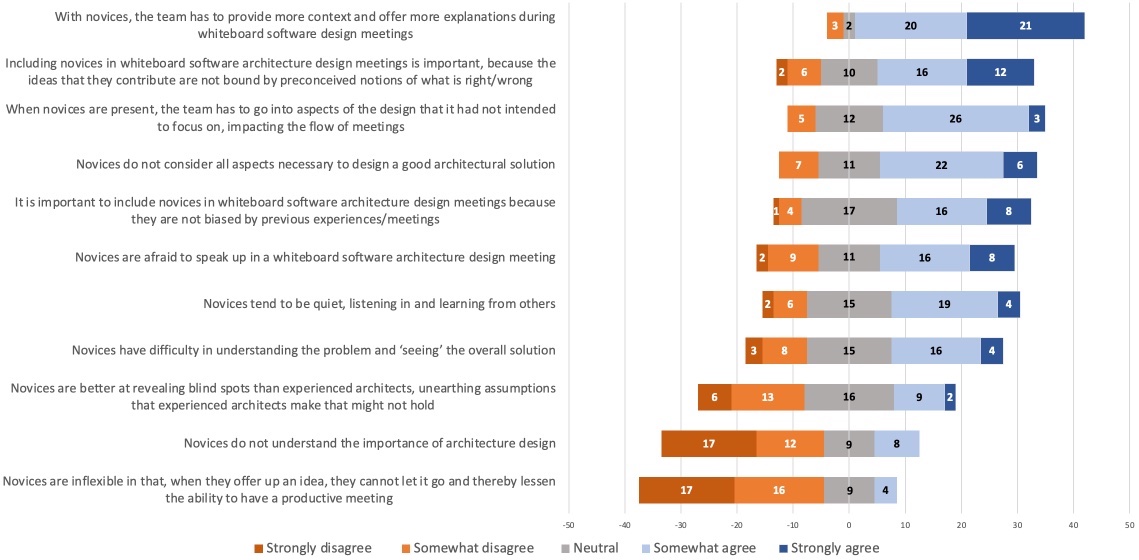}
    \caption{Architects' Perceptions on Including Novice Participants in Whiteboard Software Architecture Meetings.}
    \label{figx}
\end{figure*}

We first discuss the participation of novices. The interviewees shared eleven different perceptions on including novices in the meetings, which were ranked by the software architects who participated in the survey (Figure \ref{figx}). The top five perceptions are: \textit{with novices, the team has to provide more context and offer more explanation during whiteboard software design meetings} (average Likert score for this statement is 4.28, i.e., between "somewhat agree" and "strongly agree"), \textit{including novices in whiteboard software architecture design meetings is important, because the ideas that they contribute are not bound by preconceived notions of what is right/wrong} (3.65, between "neutral" and "somewhat agree"), \textit{when novices are present, the team has to go into aspects of the design that it had not intended to focus on, impacting the flow of meetings} (3.59, between "neutral" and "somewhat agree"), \textit{novices do not consider all aspects necessary to design a good architectural solution} (3.59, between "neutral" and "somewhat agree"), and \textit{it is important to include novices in whiteboard software architecture design meetings because they are not biased by previous experiences/meetings} (3.57, between "neutral" and "somewhat agree"). The following are some comments from the interviewees that highlight these aspects:

\xmark \textit{SA 18: "When it is with more inexperienced architects, sometimes it tends to be a bit more for lecture. Some points you end up having to go deeper to see if you bring the person to the same level or tend to go down. The person still does not make a clear division between what is architecture level and implementation. It goes down and up much more often and you have to keep pulling the person up again.}"

\cmark \textit{SA 23: "So having novices in the room who are unafraid to ask questions can help to clarify things which often leads to actual insights that would have gotten glossed over had they not been written down.}"

\xmark \textit{SA 12: "The beginner has a lot of difficulty, sometimes, to see the whole solution. The beginner is very focused on the use of technology and a little distant from the solution as a whole. This is my perception. They already want to discuss the technology, the infrastructure, the implementation, they have a very great anxiety, the more novice the developer. The tendency is to try to contain these impetus and take off the source code leading to discussion at the architecture level, regardless of technology or how it will be used. That's my main difficulty with new developers. They are still very much attached to the code and technology that will be used in the project.}"

Note that the various perceptions represent a mix of potentially positive and negative effects of including novices, so the order in which the perceptions are placed in Figure~\ref{figx} should not be read as ranging from positive perceptions (top) to negative perceptions (bottom), or vice versa.  Rather, the figure is ordered by level of agreeableness. The observation that, with novices, the team has to provide more context and offer more explanation during whiteboard software design meetings was agreed to most often, while the observation that novices are inflexible in that, when they offer up an idea, they cannot let it go and thereby lessen the ability to have a productive meeting was agreed to least often.

In our survey, we also asked respondents whether they felt that overall it is important to include novices in whiteboard software architecture design meetings. Out of 46 respondents, 23 respondents strongly agree, 19 somewhat agree, and 2 are neutral to this statement. The average Likert score for this statement is 4.35 (i.e., between "somewhat agree" and "strongly agree").

\begin{mybox}{Observation 3}
The top five perceptions on including novices in whiteboard software architecture meetings are: with novices, the team has to provide more context and offer more explanations during whiteboard software design meetings; including novices in whiteboard software architecture design meetings is important, because the ideas that they contribute are not bound by preconceived notions of what is right/wrong; when novices are present, the team has to go into aspects of the design that it had not intended to focus on, impacting the flow of meetings; novices do not consider all aspects necessary to design a good architectural solution; and It is important to include novices in whiteboard software architecture design meetings because they are not biased by previous experiences/meetings.
\end{mybox}

\begin{figure*}[h!]
    \centering
    \includegraphics[width=18cm]{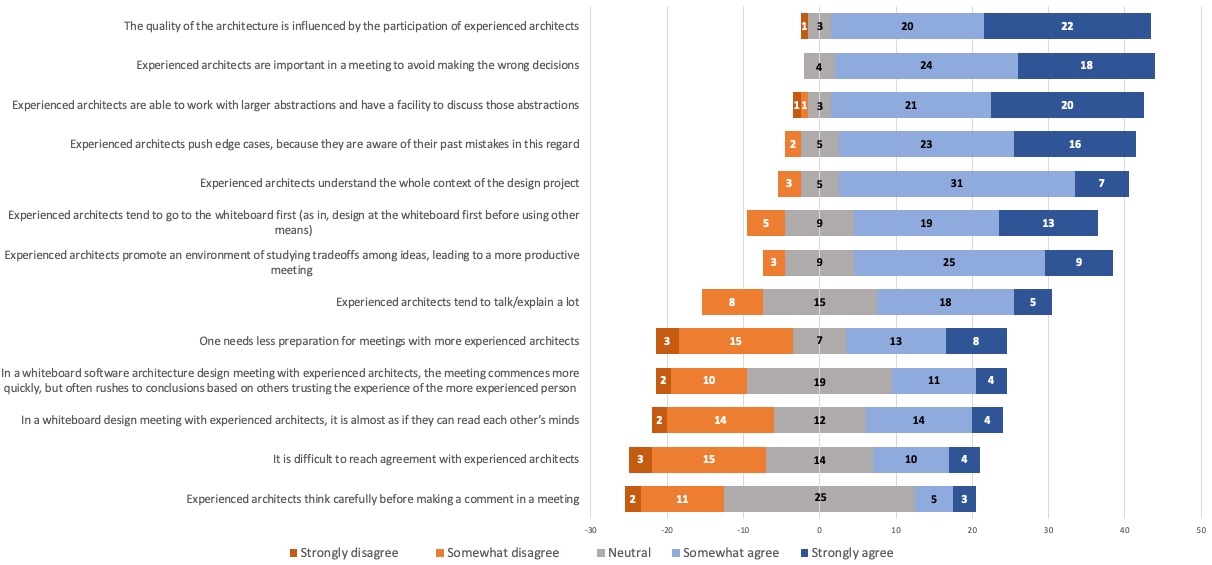}
    \caption{Architects’ Perceptions on Including Experienced Participants in Whiteboard Software Architecture Meetings.}
    \label{figurey}
\end{figure*}

In addition to building an understanding of the perceptions of architects on the participation of novices, we equally sought to understand perceptions on the participation of experienced architects in the meetings. We identified thirteen such perceptions from the interviews and asked the surveyed software architects to rank these thirteen (Figure~\ref{figurey}). Again based on the level of agreeableness, the top five perceptions are: \textit{the quality of the architecture is influenced by the participation of experienced architects} (average Likert score for this statement is 4.35, i.e., between "somewhat agree" and "strongly agree"), \textit{experienced architects are important in a meeting to avoid making the wrong decisions} (4.3, between "somewhat agree" and "strongly agree"), \textit{experienced architects are able to work with larger abstractions and have a facility to discuss those abstractions} (4.26, between "somewhat agree" and "strongly agree"), \textit{experienced architects push edge cases, because they are aware of their past mistakes in this regard} (4.15, between "somewhat agree" and "strongly agree"), and \textit{experienced architects understand the whole context of the design project} (3.91, between "neutral" and "somewhat agree"). The following are some comments from the interviewed software architects that highlight these aspects:

\cmark \textit{SA 10: "The quality of the solution with more experienced people considers requirements that less experienced people will not consider. Then we will have a more stable and robust solution with more experienced people.}"

\cmark \textit{SA 7: "It's better because sometimes some decisions take a lot of work and if the decision is made wrong by a less experienced architect setting the course of a particular project in the next three, four weeks, we will lose a lot of time. So the participation of experienced architects at these meetings is imperative.}"

\cmark \textit{SA 18: "The conversation is not the same level. You have experienced architects it is almost that you are reading one another mind. So it is more diagrammed and less talked about, and when you have a discussion it is discussions that people go deeper or that end up becoming a proof of concept because they do not have a clear answer.}"

%Figure \ref{figurey} shows the respondents’ answers with respect to experienced participants in whiteboard software architecture meetings.

In our survey, we also asked respondents whether they felt that overall it is important to include experienced architects in whiteboard software architecture design meetings. Among the 46 survey respondents, 37 respondents (80.43\%) strongly agree that it is important to include experienced architects in the meeting, 6 (13.04\%) somewhat agree, and 2 (4.35\%) are neutral. The average Likert score for this statement is 4.7 (i.e., between "somewhat agree" and "strongly agree").

\begin{mybox}{Observation 4}
The top five perceptions on including experienced architects in whiteboard software architecture meetings are: the quality of the architecture is influenced by the participation of experienced architects; experienced architects are important in a meeting to avoid making the wrong decisions; experienced architects are able to work with larger abstractions and have a facility to discuss those abstractions; experienced architects push edge cases, because they are aware of their past mistakes in this regard; and experienced architects understand the whole context of the design project.
\end{mybox}

Beyond ranking the effects on whiteboard meetings of including experienced architects, we also asked the surveyed architects to select the five most valued qualities that experienced architects should exhibit out of the eighteen identified in the discussion of Subsection 4.3.1. Table \ref{table:21} shows these eighteen qualities, as organized in frequency by which they were included in the top five. 

\begin{mybox}{Observation 5}
Posses a foundation of architectural knowledge (patterns, methods, and tools); ability to see and analyze trade-offs; experience in having designed several architectures in the past; breadth of knowledge across domains, applications, and abstractions; and ability to both introduce ideas and serve as a sparring partner for them, are the top five most valued qualities for experienced architects participating in whiteboard software architecture meetings.
\end{mybox}

\begin{figure*}[h!]
    \centering
    \includegraphics[width=18cm]{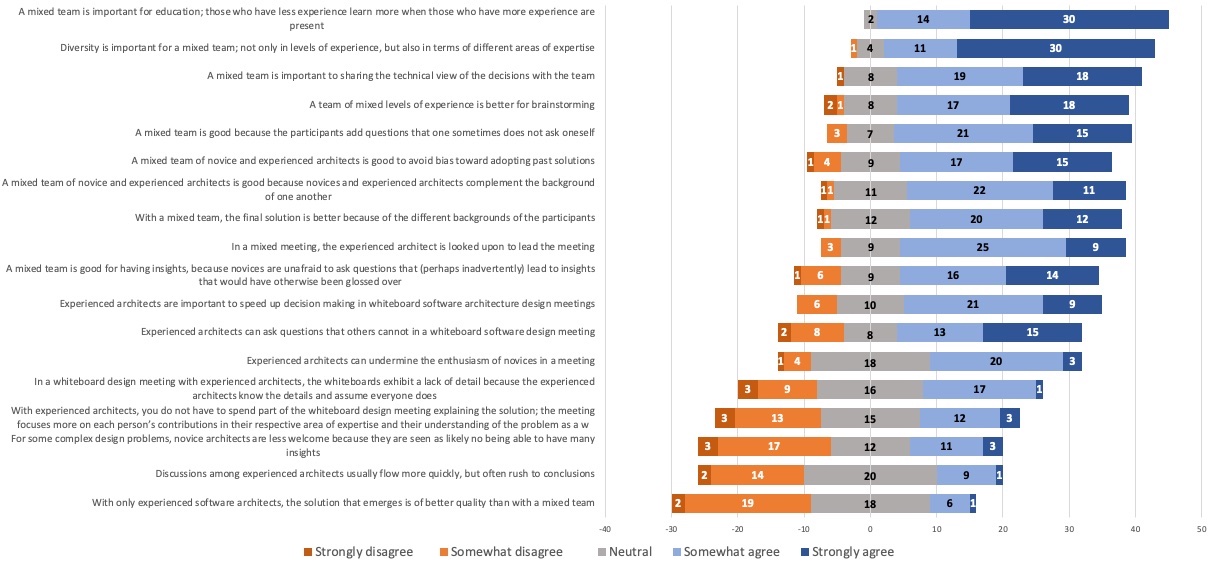}
    \caption{Architects’ Perceptions on Having a Mixed Team of Participants in Whiteboard Software Architecture Meetings.}
    \label{figurez}
\end{figure*}

Because most whiteboard software architecture meetings involve a mix of novice and experienced participants, we also asked the interviewed software architects about their perception on mixed teams, which led to eighteen different perceptions that we asked the surveyed software architects to rank (see Figure \ref{figurez}). The top five perceptions are: \textit{a mixed team is important for education; those who have less experience learn more when those who have more experience are present} (average Likert score for this statement is 4.61, i.e., between "somewhat agree" and "strongly agree"), \textit{diversity is important for a mixed team; not only in levels of experience, but also in terms of different areas of expertise} (4.52, between "somewhat agree" and "strongly agree"), \textit{a mixed team is important to sharing the technical view of the decisions with the team} (4.15, between "somewhat agree" and "strongly agree"), \textit{a team of mixed levels of experience is better for brainstorming} (4.04, between "somewhat agree" and "strongly agree"), and \textit{a mixed team is good because the participants add questions that one sometimes does not ask oneself} (4.04, between "somewhat agree" and "strongly agree"). The following are some comments from the interviewed software architects that highlight these aspects:

\cmark \textit{SA 9: "I think we should have people with various levels of experience, because those who have less experience learn more with those who have more experience. The discussion becomes more concentrated with experienced people, I think it is natural too, but it is important for everyone to participate so that everyone has the complete understanding of what is being discussed and everyone can learn how to do it, because the less experienced in the future will be those people who will lead these activities, so they need to participate from the beginning.}"

\cmark \textit{SA 11: "I do not think the result comes out better only with experienced people. I think the outcome is best suited to diversity as long as everyone is involved and participating and have confidence in each other to participate. When there is such confidence and comfort, I prefer a more diverse environment, for even in the basic question insights arise, with people who think differently, better insights emerge, so I see more diversity. Diversity not only of experience, but also of background of areas. At our whiteboard meetings, the person who is project manager, who has a different background, a business analyst, a person of operation, participates. The best insights come from this mix.}"

\cmark \textit{SA 6: "When you are working with something that does not exist and something innovative, it is interesting to have a mix of people, even because you do not know what the roadmap will be or how the architecture will materialize, how will be the implementation itself, because a lot of architectural level, you can have assertion, define interfaces among subsystems, but at the time of going into the details, mainly, for integration between existing systems, there are many details that only arise when we do a more detailed design, and this kind of question, the novices are more willing to do it for lack of knowledge and this makes you think, explain, review the concepts and you even discovering what you do not know what you thought you knew and that enriches the discussion.}"

In our survey, we also asked respondents whether it is important to include both novices and experienced software architects in whiteboard software architecture design meetings. Out of the 46 respondents, 27 strongly agree, 17 somewhat agree, and 1 was neutral with respect to this statement. The average Likert score for this statement is 4.5 (i.e., between "somewhat agree" and "strongly agree"). 

\begin{mybox}{Observation 6}
The top five perceptions on including novice and experienced participants in whiteboard software architecture meetings are: a mixed team is important for education; those who have less experience learn more when those who have more experience are present; diversity is important for a mixed team, not only in levels of experience, but also in terms of different areas of expertise; a mixed team is important to sharing the technical view of the decisions with the team; a team of mixed levels of experience is better for brainstorming; and a mixed team is good because the participants add questions that one sometimes does not ask oneself.
\end{mybox}

\begin{table}[h!]
\centering
\begin{tabular}{||p{6.3cm} c ||} 
 \hline
\textbf{ Ability} & \textbf{Frequency} \\ [0.5ex] 
 \hline\hline
 Posses a foundation of architectural knowledge (patterns, methods, and tools) & 29  \\ 
 Ability to see and analyze trade-offs & 25 \\
 Experience in having designed several architectures in the past & 21 \\
 Breadth of knowledge across domains, applications, and abstractions & 21 \\
 Ability to both introduce ideas and serve as a sparring partner for them & 19 \\ 
 Ability to communicate with the team & 19 \\ 
 Depth of knowledge in a particular area of specialty (domain, technology) & 16 \\ 
 Ability to facilitate fruitful discussions & 15 \\ 
 Ability to mentor others & 11 \\ 
 Understand failure cases & 10 \\ 
 Experience in having participated in many different projects & 10 \\ 
 Ability to stop the team from going in the wrong direction & 8 \\ 
 Understanding of the architecture of other existing systems & 8 \\ 
 Many hours of hands-on software development & 6 \\ 
 Awareness of technology trends & 5 \\ 
 Ability to interact positively with the project manager & 4 \\
 Ability to diffuse interpersonal situations & 2 \\
 Knowledge of existing design decisions and rationale & 1 \\
 \hline
\end{tabular}
\caption{Eighteen Most Valued Qualities for Experienced Architects.}
\label{table:21}
\end{table}

\subsection{Documentation (RQ3)}

Even the most suitably designed software architecture is useless if the people who need to use it do not know what it is, cannot understand it well enough to apply it, or misunderstand it and apply it incorrectly. All of the effort, analysis, hard work, and insightful design on the part of the architects will have been wasted. Thus, creating an architecture is not enough. It has to be communicated in a way that stakeholders can use it properly \cite{10.5555/1869937}. 

From the interviews, we learned that whether an architect documents their whiteboard software architecture meetings varies drastically, from never to always. We used the survey to better understand how often they do. Among the 46 survey respondents, 3 (6.53\%) always document what happens during the meetings, whereas 21 (45.65\%) document most of the time, and 11 (23.91\%) only about half of the time. Ten others (21.74\%) infrequently document what happens during the meeting and one (2.17\%) does not document ever. The following are some comments from our software architects (SA) that highlight these aspects:

\cmark \textit{SA 14: "We always take pictures of the whiteboard. Sometimes we have to delete the whiteboard, so we take a picture, then take it again, so sometimes we have three, four pictures of a single meeting. Usually this is broken. If it is a very complex flow, we digitize it in the sense of redoing that flow with some software, with some presentation or in the same flow draw itself. Sometimes we have to better specify what we have said, so we write a story and sometimes we have to break the cases and better document what each of those parts will be.}" 

\cmark \textit{SA 8: "I do not particularly like documentation very much because it is very difficult to keep the documentation current with reality. It is very common we left an architecture meeting, go to a Wiki, we draw everything, we put text, images and I have particularly never seen this evolve in conjunction with the code. You get some of reference, initial, etc, but at some point this will be outdated. At some point people will no longer update that. It's very difficult for any change people update the document to reflect what's happening. So I am particularly against documentation: over documentation.}" 

\subsubsection{Documentation Approaches}

During the interviews, we also asked the software architects to describe what forms of documentation are used. From the answers, we identified fifteen different approaches, which we seeded into the survey to understand how many of them used each of the different approaches (see Table \ref{table:31}). The top five documentation approaches are: \textit{photo(s) of the whiteboard}, \text{Wiki pages}, \textit{notes taken during the meeting by one or more participants}, \textit{notes produced/polished after the meeting}, and \textit{photo(s) with additional notes}.
%The process of creating a complex software architecture involves making several small and big decisions along the project. The results of these decisions are reflected in the views that document the architecture, i.e., the structures with their elements and relations and properties, and the interfaces and behavior of those elements, but most of the time the decisions are sadly neglected. In that case, the rationale, mainly the rationale behind the most important decisions is lost \cite{10.5555/1869937}.%
The following are some comments from our software architects (SA) that highlight these aspects, together with the documentation approaches:

\cmark \textit{SA 24: "I never document in a formal way. So, many times, what is happening is that you do is you designs things on the whiteboard and you make photos and as I said you put them either into a folder.}" [Photo (s) of the whiteboard].

\cmark \textit{SA 15: "It usually depends on the type of discussion. If I want to validate just one quick idea, just to discuss one detail, usually we only take a photo of what is on the whiteboard. This is very common. So we put the own photo of the Wiki directly, to show that it had the discussion. We try to document and to register, unless it is a type thing, we are going to join quickly and to discuss a thing of half an hour. But usually, we usually take a picture to remember what we discussed, but the general rule is to take that and write a page or two. Not documenting is an exception.}" [Wiki pages].

%Table \ref{table:31} shows the fifteen approaches used to document whiteboard software architecture meetings and its frequency.

\begin{table}[h!]
\centering
\begin{tabular}{||p{6.3cm} c ||} 
 \hline
\textbf{ Approach} & \textbf{Frequency} \\ [0.5ex] 
 \hline\hline
 Photo (s) of the whiteboard & 38  \\ 
 Wiki pages & 32 \\
 Notes taken during the meeting by one or more participants & 28 \\
 Notes produced/polished after the meeting & 17 \\
 Photo(s) with additional notes & 17 \\ 
 UML diagrams & 15 \\ 
 Powerpoint slides & 13 \\ 
 Informal record of decisions, alternatives, and rationale for choices & 11 \\ 
 Informal record of decisions & 10 \\ 
 User stories & 9 \\ 
 Flow charts & 8 \\ 
 We leave the whiteboard up (“Do not erase”) & 7 \\ 
 Architecture Decision Records (ADRs) & 6 \\ 
 Use cases & 6 \\ 
 Filling of Jira issues, stories or epics with decisions & 1 \\ 
 \hline
\end{tabular}
\caption{Fifteen Approaches Used to Document Whiteboard Software Architecture Meetings.}
\label{table:31}
\end{table}

\begin{mybox}{Observation 7}
A majority (76.09\%) of software architects always or most of the time document what happens in whiteboard software architecture meetings. Photo(s) of the whiteboard; Wiki pages; notes taken during the meeting by one or more participants; notes produced/polished after the meeting, and photo(s) with additional notes are the five most common documentation approaches used by the software architects to do so.
\end{mybox}

\subsubsection{Reasons to Document the Meetings}

Complementing whether software architects document what happens in whiteboard software architecture meetings and how they do so, we also sought to understand their primary reasons for investing time and effort in documenting. From the interview sessions, we identified twelve reasons, from which we asked each survey respondent to identify five. 

As shown in Table~\ref{table:41}, the five reasons that were most frequently selected---by a good margin---are: \textit{serve as a starting point for follow-up discussion in future meetings}; \textit{retain as evidence for later of the decisions that were made}; \textit{participants forget}; \textit{communicate the outcomes to others on the project}; and \textit{different participants have different beliefs regarding the outcomes of a meeting; documenting helps disambiguate}. The following are some comments from the software architects that exemplify some of these aspects, together with the reasons to document the meetings:

\cmark \textit{SA 13: "They are of course used because you take this as a starting point to continue the work. So, they are essential to the participants and sometimes also to other interested people who get the results by email. And later on they are looked at, as I said the meetings are just the starting point to discuss part of a solution or to provide some analysis that you can support next discussion with data.}" [Serve as a starting point for follow-up discussion in future meetings].

\cmark \textit{SA 14: "The meeting cannot stay in people's head, it needs to be documented in some way. Someone needs to commit to transferring that knowledge that has been generated, sometimes on the whiteboard itself or at other times in discussions, someone needs to be documenting this so that it becomes a task.}" [Participants forget].

\cmark \textit{SA 9: "Every time we have a new person on the team, we recommend [ed: them] to see the architecture documents and present doubts or explain his understanding of the architecture. It is to make sure [ed: they] understood what we designed. So, the documents are used as reference, whenever someone has a doubt about a piece or a component designed, we recommend the person to go back to the architecture document.}" [Use later to educate new people on the project].

\begin{table}[h!]
\centering
\begin{tabular}{||p{6.3cm} c ||} 
 \hline
\textbf{Reason} & \textbf{Frequency} \\ [0.5ex] 
 \hline\hline
 Serve as a starting point for follow-up discussion in future meetings & 38  \\ 
 Retain as evidence for later of the decisions that were made & 33 \\
 Participants forget & 30 \\
 Communicate the outcomes to others on the project & 27 \\
 Different participants have different beliefs regarding the outcomes of a meeting; documenting helps disambiguate & 26 \\
 Use later to educate new people on the project & 14 \\ 
 Validate in detail whether the design ideas indeed can work as intended & 11 \\ 
 Train the team on the design & 8 \\ 
 Enable reuse of the design ideas in other design projects & 8 \\ 
 Present a preliminary solution to the customer & 5 \\ 
 Participants sometimes second-guess what they did & 5 \\ 
 Include as part of the design that we are handing off to the customer (so they can do the implementation work) & 3 \\ 
 
 \hline
\end{tabular}
\caption{Twelve Reasons Why Software Architects Choose to Document Whiteboard Software Architecture Meetings.}
\label{table:41}
\end{table}

\begin{mybox}{Observation 8}
The top five reasons to document whiteboard software architecture meetings are: serve as a starting point for follow-up discussion in future meetings; retain as evidence for later of the decisions that were made; participants forget what was discussed; communicate the outcomes to others on the project; and different participants have different beliefs regarding the outcomes of a meeting/documenting helps disambiguate.
\end{mybox}

\subsection{From Whiteboard Architecture to Code (RQ4)}

%The same way as developers \cite{Cherubini:2007:LGW:1240624.1240714} and software designers \cite{Mangano:2014:SID:2556288.2557411}, software architects create sketches during whiteboard software architecture meetings to understand, to design, and to communicate their solutions. 

%Previous work has investigated the lifecycles of sketches \cite{6069462, Mangano:2014:SID:2556288.2557411, 8091190} and source code artifacts which sketches are related \cite{10.1145/2635868.2635891}. However, the transition from sketches created at whiteboard to code and their reasons to change are important aspects not explored. 

Documenting the outcomes emerging from a whiteboard software architecture meeting is important, but equally important is recognizing that what initially has been designed and documented may well change. Such changes may actually already emerge just from the act of documenting (an architect more formally documenting decisions after a meeting may realize that some aspect of the architecture as designed has a flaw and decide to address the flaw on the spot since it is not too difficult) or may happen later (for example, when the architect debriefs the team on the decisions made and in the process of explaining realizes an issue, or when a developer responsible for making some changes in the midst of making those changes encounters a problem in how the envisioned changes are incompatible with some aspect of the current code).  While much has been said to date about the problem of architectural decay \cite{DBLP:journals/smr/LiLSA22} and erosion \cite{DBLP:journals/jss/GurpB02}, little work to date has studied what kind of decay and erosion happen (exceptions exist, e.g., \cite{DBLP:journals/ese/AliBOHB18}) and no work to date has examined the deviations from whiteboard software architecture meetings to eventual code.

In our interviews, we asked architects about what kinds of changes they have witnessed being made to the architectural designs and plans they have created in whiteboard software architecture meetings.  Table \ref{table:5} shows the ten different aspects that were recounted by the software architects, in order of how often they were experienced by the surveyed software architects. The top five aspects that change from the whiteboard to implementation are: \textit{interface of major components in the architecture}; \textit{implementation details}; \textit{a small handful of components in the architecture}; \textit{detailed modules inside architectural components}; and \textit{database schema}. The following are some comments from the interviewed software architects that highlight these aspects, together with the kinds of changes:

\cmark \textit{SA 9: "I think when you do not think of all the details. I remember that we were once designing part of the KNoT device protocol and we thought and wrote it on the whiteboard, but we did not think at all. So we had to add more things we had not thought about when we were designing. They are things like that. They are parts of the implementation that you do not think about when you're on the whiteboard, but then you realize you need it in implementation. It is which I think usually changes.}" [Implementation details].

\cmark \textit{SA 8: "What I see to change is usually the format of the messages. We define the messages with certain attributes and then we see that something is missing and we have to adjust with more or less attributes due to performance, for example, the payload got too big and we need to decrease the payload due to performance or we need to send more attributes because the domain or functional requirement changed on the other side and we will need to consume more information than what was originally intended. It is something that usually changes a lot over the life of an application. It is something I see happen frequently, we define that the message will be this and for some reason we have to change that message or increase or decrease for some reason.}" [Format of messages connecting various parts of the architecture].

\begin{table}[h!]
\centering
\begin{tabular}{||p{6.3cm} c ||} 
 \hline
\textbf{ Aspect} & \textbf{Frequency} \\ [0.5ex] 
 \hline\hline
 Interface of major components in the architecture & 34  \\ 
 Implementation details & 28 \\
 A small handful of components in the architecture & 24 \\
 Detailed modules inside architectural components & 21 \\
 Database schema & 20 \\
 Driving scenarios & 19 \\ 
 Format of messages connecting various parts of the architecture & 13 \\ 
 Overall architectural solution & 13 \\ 
 Key algorithms & 7 \\ 
 Change of proposed implementation (time or scope) & 1 \\
 
 \hline
\end{tabular}
\caption{Ten aspects that change from sketch to implementation.}
\label{table:5}
\end{table}

\begin{mybox}{Observation 9}
The top five aspects that change from whiteboard architecture to implementation are: Interface of major components in the architecture; Implementation details; A small handful of components in the architecture; Detailed modules inside architectural components; and Database schema.
\end{mybox}

\subsubsection{Rationale for Changes}

After identifying the aspects that change from sketches to code, a natural step was to understand the rationale for the changes. Using a similar approach, from the interviews with the software architects we identified seventeen reasons to make changes from whiteboard architecture to implementation and then asked the surveyed software architect to mark the five that they experienced most frequently. Table \ref{table:6} shows the seventeen reasons to make changes together with their frequency. The top five reasons to change are: \textit{certain aspects of the solution are over-simplified and turn out to be more complex}; \textit{architects discover a better solution than the original at the whiteboard}; \textit{multi-dimensionality of the problem – qualities that were not considered (or merely lightly considered) during the whiteboard meeting are negatively affected by the planned solution}; \textit{the project is Agile, and thus had to respond to new circumstances}; and \textit{performance}. 
The following are some excerpts from the interviews that highlight a pair of these aspects, together with the rationale for the changes:

\cmark \textit{SA 19: "Good question. What sometimes changes, in fact, does not change so much because what we do on the whiteboard is very macro, it is an overview, it does not go as far as detail, but what sometimes changes is that sometimes you think of using a technology, a framework, and when you go to Google, Stack overflow, or talking to a colleague, you discover another technology is better. For example, a text search with Solr, and you find out that everyone is using Elastic Search, and you think: let's use Elastic Search. So you find you have a more interesting alternative.}" [We discovered a better solution than the original we devised at the whiteboard].

\cmark \textit{SA 13: "So, sometimes you get to know more details about some aspects so, we have multi-dimensional problems usually. So, this is embedded software which is variable, being embedded it is also resource-constrained, it is real-time running on the multicore processors and there are also some other architectural qualities, maintainability and so on that need to be addressed. So, frequently the whiteboards discussions concentrate on some specific quality or on some specific problems. Only then you realize: Ok if we do it that way then maybe a third or fourth architectural quality will suffer.}" [Multi-dimensionality of the problem – qualities that were not considered (or merely lightly considered) during the whiteboard meeting are negatively affected by the planned solution].

\begin{table}[h!]
\centering
\begin{tabular}{||p{6.3cm} c ||} 
 \hline
\textbf{ Reason} & \textbf{Frequency} \\ [0.5ex] 
 \hline\hline
 Certain aspects of the solution were over-simplified and turn out to be more complex & 28 \\
 We discovered a better solution than the original we devised at the whiteboard & 23 \\
 Multi-dimensionality of the problem – qualities that were not considered (or merely lightly considered) during the whiteboard meeting are negatively affected by the planned solution & 23 \\
 The project is Agile, and thus had to respond to new circumstances & 23 \\
 Performance & 19 \\ 
 Customer requirements changed midstream & 17 \\ 
 Technology/platform limitations & 17 \\ 
 Team made false assumptions & 14 \\ 
 Difficulty in mapping the high-level solution to actual code & 9 \\
 Scalability & 9 \\
 Certain predictions of how the architecture would behave did not hold up & 8 \\
 Reliability & 7 \\
 Lack of having documented what we did at the whiteboard & 3 \\
 Team misunderstood the architectural design & 3 \\
 Social problems with the team & 2 \\
 All the above at different times, it is very context dependent & 1 \\
 The original meeting was not conducted very well and thus not effective & 1\\
 
 \hline
\end{tabular}
\caption{Seventeen Reasons to Make Changes from Whiteboard Architecture to Code.}
\label{table:6}
\end{table}

\begin{mybox}{Observation 10}
The top five reasons to make changes from sketches to implementation are: certain aspects of the solution are over-simplified and turns out to be more complex; architects discover a better solution than the original at the whiteboard; multi-dimensionality of the problem – qualities that were not considered (or merely lightly considered) during the whiteboard meeting are negatively affected by the planned solution; the project is Agile, and thus had to respond to new circumstances; and performance.
\end{mybox}

\subsubsection{Missing Meeting Aspects as Potential Causes}

One somewhat unexpected theme that emerged from the interviews is that the architects talked about `what could have been': aspects of whiteboard software architecture meetings and how they were conducted that, had they been done differently, could perhaps have avoided future changes being necessary. Based on the interviews, we identified sixteen such aspects from which, once again, each surveyed software architect could tag five as 'missing aspects': aspects that had they been incorporated better may have improved prior meetings. 

The top five aspects that resulted were: \textit{sufficient information about the problem to design the solution}; \textit{understanding of the relative priority of various design considerations}; \textit{metrics that delineate ‘success’ of the architectural design}; \textit{validity of assumptions about decisions, as to whether they hold up at implementation time}; and \textit{details about the envisioned implementation}. Table \ref{table:7} shows the sixteen aspects in order of frequency in which they were tagged by the surveyed architects.
The following are some comments from the software architects:

\xmark \textit{SA 12: "That's a good question. I think it's the depth of the impact of architecture on the solution. Sometimes this happens, for example, we integrate a video platform of a television channel with 110 thousand videos approximately. We drew the entire model while we did not have the total volume of videos from the customer. Sometimes we have little knowledge of the whole context the solution requires and this has a lot of impact on implementation. It makes a very big difference you implement a solution for 10,000 and another for 110 thousand videos with 3T of storage. What is missing sometimes on the whiteboard is enough information to get the solution.}" [Sufficient information about the problem to design the solution].

\xmark \textit{SA 8: "Success criteria and evaluation criteria. Whether it is good or not and how we are going to measure things. As we measure performance, uptime, fault tolerance, this kind of thing is invariably missing.}" [Metrics that delineate ‘success’ of the architectural design].

\begin{table}[h!]
\centering
\begin{tabular}{||p{6.3cm} c ||} 
 \hline
\textbf{ Aspect missing} & \textbf{Frequency} \\ [0.5ex] 
 \hline\hline
 Sufficient information about the problem to design the solution & 29 \\
 Understanding of the relative priority of various design considerations & 24 \\
 Metrics that delineate ‘success’ of the architectural design & 19 \\
 Validity of assumptions about decisions, as to whether they hold up at implementation time & 19 \\
 Details about the envisioned implementation & 19 \\ 
 Certain requirements & 18 \\ 
 Agenda for the meeting & 14 \\ 
 Details about the current implementation & 13 \\ 
 Test cases governing the architectural design & 9 \\
 Dependencies among the various whiteboard sketches & 9 \\
 Context diagram & 9 \\
 Interfaces among the components & 8 \\
 Structure of the messages exchanged by the components & 2 \\
 An overview of the project & 2 \\
 Clear problem to be solved & 1 \\
 Clear next steps and assigned responsibilities & 1 \\

\hline
\end{tabular}
\caption{Sixteen Aspects Missing from Whiteboard Software Architecture Meetings that Could Have Improved the Outcomes.}
\label{table:7}
\end{table}

Note how these factors represent a mix: some concern having additional information at hand, some the conduct of the meeting itself, some additional angles of the design that they wished they had worked out in more detail, and some the criteria by which the architecture eventually would be judged. 

\begin{mybox}{Observation 11}
The top five aspects missing from the whiteboard software architecture discussions are: sufficient information about the problem to design the solution; understanding of the relative priority of various design considerations; metrics that delineate ‘success’ of the architectural design; validity of assumptions about decisions, as to whether they hold up at implementation time; and details about the envisioned implementation.
\end{mybox}

\subsection{Digital Tools (RQ5)}

Previous research has shown that different kinds of media are used for architecture design.  Beyond the whiteboards, these media may include scrap paper to informally sketch and model, but also software tools like Photoshop and Powerpoint~\cite{10.1145/2635868.2661672}. The past decade years also has seen the emergence of a new crop of tools, such as the Microsoft Surface and other devices which enable touch based design and cloud-based, remote collaboration oriented whiteboard tools such as Gliffy\footnote{https://www.gliffy.com/} and Miro\footnote{https://miro.com/}. These kinds of tools offer new opportunities, both in terms of how tams work together and who is brought into meetings (e.g., remote participation is much easier so meetings can be more inclusive) and in terms of moving outcomes downstreams (e.g., many of these tools have export capabilities, some are tightly integrated with other tools such as Wikis and task managers). Thus, it is important to understand software architects' perceptions about these digital tools and the impact on their activities. We used the survey to do so.  

Among the 46 survey respondents in our survey, 21 (45.7\%) software architects already had experience in using a digital whiteboard tool in their software architecture design meetings. Twenty-five (54.3\%), however, had never used such a tool. We asked the participants about whether they currently still use a digital whiteboard tool in the meetings, with 10 (21.7\%) of the participants actively using such a tool and 36 (78.3\%) not (even if they had experimented with these kinds of tools before). Finally, we asked the participants whether they would like to use a digital whiteboard tool in a software architecture design meeting: 34 (73.9\%) participants said that they would like to use such tool and 12 (26.1\%) preferred not to. The following are some comments in this regard:

\cmark \textit{SA 7: "Perhaps, one thing I have great curiosity about being used are the digital whiteboard meetings that store those artifacts in some media shared among all team members. I do not remember seeing this working on any team. In fact, I've never been involved in any project that has used something like Microsoft Surface. Something you scribbled on the spot and everyone could record that to be shared later. To be kept as a whiteboard record without being a photo, something that you could search over, something that you could consult very fast without being an image basically. Perhaps this changed the performance of architectural meetings or perhaps facilitated employee turnover, entering or leaving a new developer or team developer, he could gain access to architectural evaluations history, architectural meeting history, and everything that was written in that digital whiteboard. I wanted to have this notion, but I have not, I do not know a team that has used this kind of digital artifact to make life easier. I'm more curious actually.}"

\xmark \textit{SA 28: "The idea of using digital whiteboards is a good one, but I've never done it. I guess because the current solutions available neither offer satisfactory latency nor they provide the natural and seamless experience that make whiteboards meetings appealing in the first place.}"

\begin{mybox}{Observation 12}
45.7 \% of software architects already used a digital whiteboard tool in a software architecture meeting. However, only 21.7\% of them are currently using a digital whiteboard tool in their projects. 
\end{mybox}

\section{Implications}

Our study takes a look at software architecture meetings at the whiteboard, with its findings representing a first set of observations concerning these meetings. Some of the findings align with findings that have been made about whiteboard meetings more broadly. As one example, the fact that architects go to the whiteboard for a variety of reasons aligns with programmers using the whiteboard for many different reasons~\cite{Cherubini:2007:LGW:1240624.1240714}. Other findings align with the literature on expertise and the roles that experts and novices play in creative endeavors by, for instance, confirming that the questions asked by novices can cause experts to reflect more deeply and as a result reassess assumptions that hitherto had not been considered important to discuss~\cite{LEE2022101089}. Our paper anchors these general findings into the specifics of whiteboard software architecture meetings and combines them with other, original findings pertinent to this setting. Below, we discuss the collective implications for research, practice, and tools.

\textbf{Research.} Different meetings will feature a different mix of levels of experience in participants (e.g., mostly novices, a relatively even mix of experienced architects and novices, mostly experienced, exclusively experienced). Some of this will depend on the purpose of the meeting as well as the stage of development of the architecture. The perceptions of the architects as discussed in Section~\ref{subsec:experience} on what they believe the effects of different levels of experience are on how the meetings proceed remain perceptions. The fact that many of the perceptions receive a significant amount of agreement from the surveyed architects implies it is likely that many of these perceptions are largely accurate. At the same time, it is important to verify these perceptions with rigorous studies on the impact of the mix of experience in team composition on both how whiteboard software architecture meetings proceed as well as their eventual outcomes. For instance, one of the software architects observed: \textit{"The quality of the solution with more experienced people considers requirements that less experienced people will not consider. Then we will have a more stable and robust solution with more experienced people.}" Yet, as we already mentioned, it is also believed that novices can cause experienced architects to reconsider aspects of their design because of seemingly `ignorant' questions, which equally can impact the resulting quality. Exactly where the balance lies will need to be studied carefully, perhaps along the lines of the experiments of ~\cite{REIMLINGER2019204, GE2021101020, SILK2021101015, KRISHNAKUMAR2022101133} .

The behavioral and psychological aspects of these kinds of meetings should also be further investigated. Our findings are varied in this regard, ranging from the perception that novice participants may be "afraid to speak", to the "difficulty to reach agreement" with merely experienced people, to the impact of "inflexible participants" and "dominant know-it-all personalities", to the importance of good facilitation to having a successful meeting. \textit{"Another advice has to do with facilitation as well. It is making sure everyone in the meeting is heard. Sometimes we have more talkative and less talkative people and sometimes we have opinions that are left out because some people are more shy or not so vocal. So to have an effective meeting, facilitation is a crucial point.}" Studies examining meeting conduct and people interactions exist in a more general sense (e.g., ~\cite{doi:10.1080/01446190701567413, PALETZ201739}), but the domain of software and particularly software architecture has not been studied to date in that regard. With software exhibiting unique characteristics and challenges when it comes to meetings at the whiteboard, observational studies considering these aspects are welcome. 

The connection from whiteboard software architecture meetings to eventual code should also be further investigated. As one architect commented: \textit{"It depends a lot on the context. But the ones I've worked with most intensely, they were performance decisions. For example, some specific strategy of data processing. Which strategies were going to be used. It was fully reflected in the code. Because we drew the threads, how many threads would be used, what size the data chunk would be considered, who would play the role of the reader, who was going to turn the data, and who was actually going to save, so in that case it was reflected directly in the code. Not all of them are, but these and others related to performance problems, which I remember and participated in, all reflected in the code.}" Such detailed traceability of in this case architectural design decisions is rarely the case. Indeed, from our findings, it is clear that the architects are keenly aware a disconnect exists, that a range of canonical changes tend to be necessary when an architecture as designed is refined into actual code, and that proper ways of documenting what happens at the whiteboard to inform future development remain lacking. The use of Architectural Decision Records (ADRs) has recently gained some traction (six out of 46 surveyed said they have used ADRs, conform Table~\ref{table:31} and the literature is also reporting on the role of ADRs~\cite{DBLP:conf/zeus/KoppAZ18}), with one of the architects commenting: \textit{"Lately, we are experimenting with a technique called ADR, a template that we put in archives of the repository we are developing that documents the decisions. Then you open a pr file, with that decision, someone approves immediately and gets that decision. In general, it is a very simple and short document, sometimes it does not reach half a page of a document, but we record a decision that we want to record and return to it when we are making other decisions.}" How the use of ADRs influences the ability to better connect whiteboards meetings to code is unknown and ADRs are merely one of multiple possible approaches. While studies exist of architectural decay and erosion in the literature (e.g, \cite{DBLP:journals/jss/GurpB02, DBLP:journals/ese/AliBOHB18,  DBLP:journals/smr/LiLSA22}), these studies tend to compare the as designed architecture with the as implemented architecture. Exactly how and why decay and erosion took place over the life of a system, however, is not documented (i.e., the various reasons from Table~\ref{table:6} mapped onto actual moments and context within the development project). Field studies that examine precisely where and when breakdowns occur, how teams overcome those breakdowns, and what tools were in use yet failed to provide the necessary support are necessary. 

%Other important aspects which should be investigated in future studies include:\textit{documentation}, \textit{design decisions}, and \textit{distributed meetings}. 
%Documentation is a controversial topic in software engineering ranging from source code documentation to software architecture. 

%Some design decisions are preserved from sketches at whiteboard to code. On the other hand, other decisions are not preserved. The first case was reported by one software architect discussing performance: \textit{"It depends a lot on the context. But the ones I've worked with most intensely, they were performance decisions. For example, some specific strategy of data processing. Which strategies were going to be used. It was fully reflected in the code. Because we drew the threads, how many threads would be used, what size the data chunk would be considered, who would play the role of the reader, who was going to turn the data, and who was actually going to save, so in that case it was reflected directly in the code. Not all of them are, but these and others related to performance problems, which I remember and participated in, all reflected in the code.}" Studies to better understand the life cycle of sketches and the relationship from sketches at whiteboard to code are also important. 

We also note that, while our study was agnostic whether the whiteboard meetings were entirely collocated, fully remote, or hybrid, we note that many architects imagined and recited collocated meetings in their answers. While we did ask the interviewed and surveyed architects about their use of and interest in electronic whiteboard tools to support distributed participation, such use is still limited and it is therefore important to in future more explicitly focus on the needs in distributed and hybrid settings, particularly given today's COVID-19 induced reality but certainly not exclusively so as distributed work equally took place before COVID-19: \textit{"We've got hugely distributed teams so face to face meetings are becoming less and less important. In fact, I've been working almost entirely remotely with people so we kind of we do sometimes have we can make a virtual whiteboard at design meetings but you know I think the tools for that are fairly primitive now, right.}" A great many studies are emerging at this time surrounding the topic of remote and hybrid meetings (e.g.,~\cite{DBLP:conf/chi/DAngeloG18, DBLP:journals/tosem/FordSZBJMBHN22}), including some emerging work on maintenance design by an architecture team~\cite{DBLP:conf/icse-chase/SoriaHB22}, yet a focus on the creative and design aspects of whiteboard software architecture meetings remains absent.

%Empirical studies on distributed meetings are important to better understand the limitations of this environment and its impact on the software architecture, as well as identifying new features for software architecture tools. 

% Understanding different levels of experience, for example, it is very important since much of our knowledge of software engineering expertise come from studying new engineers rather than experienced ones \cite{7194618}.

Finally, we note that the capture of information from these meetings is considered important~\cite{DBLP:journals/jss/CapillaJTAB16}, but little to no studies exist of how their importance actually plays out downstream in practice. Intuitively, the architects know they need to capture what was being discussed, but scant literature exists that shows the benefits of doing so: when is this info used, by whom, how is it making a difference, and what happens when the information is not available? Such studies could change improve the understanding as to why one should capture outcomes from the meetings in a concrete form.

\textbf{Practice.} 
From our study, several important suggestions arise for practicing software architects and how they choose to conduct and engage in whiteboard software architecture meetings. First and foremost is the consideration that architects should carefully select the right mix of participants to the meeting.  While this sounds in some ways is too straightforward and perhaps even redundant advice, as architects typically do consider whom they invite to the meetings and why, three dimensions stand out to which they should pay particular attention: experience, different perspectives, and relevant expertise.  In terms of experience, the architects that we studied strongly feel that mixed levels of experience should be brought into the room, from highly experienced architects to much more novice architects. Each group challenges the other, causing broader discussions to take place that both consider aspects of the architecture that otherwise would not be considered and teach the novices how to become better architects through their participation (a key trait of experienced software designers is continuous learning about new technologies and other types of systems as key traits).

Beyond such mixed levels of experience, including meeting participants who bring different perspectives to the discussion is crucial: \textit{"What also helps when you have 2-4 participants is that they should have different perspectives, so they come from different organizational background or have a different expertise focus. You cannot know everything as a single person and if you get the second person which is similar to you from the profile it does not double the knowledge. But if you have a few people with different perspectives then you shed light from different directions on the problem and usually someone has a different perspective and sees other aspects of the problem which you could not come with because you do not even know that such thing exists".}  Complementing experience and perspective is the importance of including people who have the relevant expertise: \textit{"Another aspect is to bring the right people to the meeting. I have seen meetings that were not effective because we did not have the right people at the meeting. Let's discuss deploy, containerization, but no one knows enough of Docker to talk about it, does not know what the possibilities are, etc., so bringing people who know how to talk about it is important to get the findings faster".} It still happens all too frequently that meetings are conducted that fail these inclusion criteria.

Hand-in-hand with whom to bring into the meeting is the fact that architects should promote psychological safety for the participants. \textit{"Promote psychological safety, that is, psychological security for people to express opinions, so they do not feel frightened. When you are going to make a comment, which is a complete bullshit, that's fine, this should not have a consequence, it should not be mocking an opinion of a person who is sincere and is willing to contribute to the meeting. So the person leading the meeting needs to worry about all of these aspects so he can extract the most value from it".} The importance of such psychological safety is well-known in the literature on how to conduct high-quality meetings (e.g., ~\cite{delizonna2017high, NEWMAN2017521}), but it is an important reminder for architects to recognize that one of their roles in these meetings is to create a welcoming and open environment for discussion. 

Other well-known strategies for conducting high-quality meeting were recognized by the architects, ranging from  making sure that everyone is heard 
(\textit{"Sometimes we have opinions that are left out because some people are more shy or not so vocal. So to have an effective meeting, facilitation is a crucial point".}), 
defining and publicizing an agenda well before the meeting
%Other "basic aspects", such as define an agenda for the meeting, share some material before the meeting, and summarize the meeting in the end were also commented. As explained by one of the software architects: 
(\textit{"I think that it is very important to have a meeting agenda. Sometimes it happens the meeting gets away from the topic and what we do is to set another meeting for the new, another topic. We try to stay focused on the problem that we have on our hands and use the time exactly for that".}),
to sharing relevant materials beforehand so that meeting time can be spent constructively considering materials that have been read by the participants before the meeting starts rather than actually reading the materials on the spot, to 
%Regarding the productivity of the meetings, people should "share some material in advance" if they see that the understanding during the meeting will take a long time to be understood; "limit the number of participants" to no more than six people; "include experts" to go quickly for an initial solution; and 
including a "facilitator" (\textit{"Regarding agreeing and making the meeting effective, it has a bit of facilitation as well. If facilitation is active, we can reach conclusions faster. Usually we are discussing a diagram or some proposals and we have opposing opinions, different proposals, and sometimes the quickest conclusion is: let's test both. And there must be a maturity in the facilitation to reach that consensus quickly."}). In many ways, whiteboard software architecture meetings are just another type of meeting, so it is not surprising that these kinds of general lessons also apply here. We do note the importance of sharing relevant materials beforehand. From Table~\ref{table:7}, it is clear that a significant problem in these meetings is a lack of critical information, with the top four being insufficient information of the problem, knowledge of the relatively priority of various design considerations, metrics that delineate success, and an understanding of the assumptions being made and how valid those assumptions are. This is information that an architect could and should prepare beforehand, as it otherwise can become quickly lost in the discussion or, more likely yet, be seen in the meeting as a distraction from progress when participants choose to bring up not knowing something and asking for clarification, or, worse yet, simply be forgotten to be brought into the discussion.

The role of the architect themselves is also important in setting an example for others to follow.  For instance, experienced software architects must defend their decisions but not be hostage to them: \textit{"You can discuss and understand the arguments, defend your proposal and also have the discernment, although difficult, that if another alternative presented is better than yours, you can also understand that it is the one that should be chosen".} Similarly, experienced architects typically set an example in 
%"Involve less experienced people is also important": \textit{"So I would say in general, a better practice from my perspective would be also to involve less experienced people because then they develop in the long term. That is also what we try to do partially because of that and partially because of the fact is not possible to experts to do all the work because they are not so many experienced architects."} 
explicitly seeking out to think about and discuss corner cases \cite{DBLP:journals/software/ErderP17}: \textit{"Because sometimes we are just thinking about the happy flow and what are the possible shits that will happen? So having a person who is going to be provoking this, provoking in a way that everyone can hear the opinion of each other, I think it is a very well conducted meeting. Any meeting that is very partial, only one person talking too much is a meeting that is not being well conducted."} In so setting the tone, the architect invites perspectives and ideas and also shows how to examine the emerging design or architectural changes from all angles, without necessarily being negative about it. After all, it is a work in progress and its eventual quality is essential.

Documentation is something that must be considered carefully. On the one hand, the meeting cannot stay in people's heads. On the other hand, it is important to not spend too much effort documenting: \textit{"Because it will change and it is very difficult to maintain the consistency of what is in the code with this abstraction which is its architecture. It is to spend bullet with deceased dead. It should be used as a reference, as I said: you say when we started was like that and we changed because of these aspects, it is much more static you document the principles that were used in your architecture than the architecture draw."} That said, it is clear that the interviewed and surveyed architects firmly believe some form of documentation is needed, but still struggle with the best ways of doing so. In terms of concrete advice, it might be worthwhile to focus the documentation aspects on those parts of the architecture that are likely to change (conform Table~\ref{table:5}). This is somewhat a counter-intuitive idea, as normally one tends to concentrate on documenting those parts of the architecture that are well understood and firm.  Yet, documenting those parts that are more likely to change has the potential benefit of creating artifacts that can be discussed earlier (and thus with less potential cost in terms of already implemented code) and that, by virtue of perhaps being documented as tentative can invite such further discussion.  Moreover, when changes are needed, they can be done with an explicit representation of what was discussed in the past, which represents an important starting point and avoids having to re-invent the wheel or re-constructing the prior discussion.

%"Document the architecture in high level", "be consistent with the notation defined in the whole documentation", and "use a notation which people know" are also important aspects. According to one architect, \textit{"Perhaps if we could design and make available the digital artifact of the architecture in the most high level possible way, by not defining classes, methods, everything in detail, but simply the large modules and large components of the system that are projected on the wall or in a slide to everyone looks at and knows how a certain component works, how it sends messages to another and returns something, and it takes so much time to answer, etc. This macro view of the system is a great advice".}

\textbf{Tools.} Beyond the traditional physical whiteboard with pens and an eraser, which still continues to be used often for in-person meetings, many tools have been developed that provide a virtual whiteboard experience, enabling team of remote participants to work together (e.g., Miro\footnote{https://miro.com/}, Jamboard\footnote{https://workspace.google.com/products/jamboard/}, Mural\footnote{https://www.mural.co/}, ConceptBoard\footnote{https://conceptboard.com/}). Particularly over the past few years, these tools have become increasingly popular and have seen a significant expansion in the types of features they include.  That said, from the interviews and the survey, room for improvement exists.  One aspect concerns the relationships among whiteboard sketches. When they work at the whiteboard, whether in a single meeting or across a series of meetings, architects tend to produce a wide range of sketches~\cite{DBLP:journals/tse/ManganoLPH15}. These sketches relate to one another in all sorts of ways, be it one sketch being a refinement of another, a sketch offering an alternative to another sketch, or some sketch providing a UI that is tied closely to the architecture being worked on in another sketch, among others.  Such relationships can help both in understanding the sketches at some later point in time as well as in organizing them proactively for later usage.  \textit{"Because you always try to be fast so to say and to focus on certain aspects. But also, as a whiteboard does not really allow for it you don't go back and look at a series over maybe five to 10 sketches aligned with each other again. And as you're not really doing that and you're not really creating consistency, this is often missing in the end."}  The kind of functionality being explored in Calico to explicitly type the relationship between different whiteboards may provide a starting point for providing this kind of functionality~\cite{DBLP:conf/kbse/ManganoBDNH10}.

Another aspect that was mentioned frequently was collaborative design.  Nearly every virtual whiteboard already supports multiple participants working in parallel on the same sketch, meaning that, on the one hand, collaborative design is already supported.  On the other hand, as one software architect said:"\textit{So there's a lot of challenges for that. It's time zones, as well as well as the technology that lets us do collaborative design online. So I think that a lot more research needs to be done into how to do distributed collaborative designs. How people can work together when they're not physically co-located. We have a lot of it, if we're timezone compatible like I do a lot of work with people. So, we're only an hour off and we have communication mechanisms you know instant messaging mechanisms of various kinds of let up but let us ping somebody. Kind of like just walking down to the cube and talking to them. But then to do it to actually do any kind of design it usually generates just into a voice call it with no other tools, so it gets to be kind of hard to manage."} What this comment points to is a mode of work that blends synchronous and asynchronous participation. Existing tools are ill-equipped for such work, focusing on support for synchronous collaboration.  That said, features such as commenting through sticky notes do begin to move in the right direction, but more is needed, both in terms of actually understanding what the needs of architects are when it comes to such blended group work and how to best support those needs.

Search and retrieval features were also mentioned. \textit{"Perhaps, one thing I have great curiosity about being used are the digital whiteboard meetings that store those artifacts in some media shared among all team members. I do not remember seeing this working on any team. Something you scribbled on the spot and everyone could record that to be shared later. To be kept as a whiteboard record without being a photo, something that you could search over, something that you could consult very fast without being an image basically."} While existing virtual whiteboards do support keeping whiteboards and their sketches around in perpetuity, and while existing features such as folders and tagging help, the more sketches are kept, the more difficult it becomes to locate that one sketch one might need. Research has studied tagging as a means of retrieving video segments (e.g.,~\cite{DBLP:journals/mta/BartoliniPR13}), handwriting recognition so to beautify and make machine readable---and thus searchable---handwritten text in sketches (e.g.,~\cite{DBLP:conf/chi/WhittakerHW94, DBLP:conf/bcshci/PlimmerF07}), and voice-based snapshots to capture and make searchable important design moments (e.g., ~\cite{DBLP:conf/chi/StifelmanAS01, DBLP:conf/icse/SoriaH19}), all of which could provide inspiration and a technical basis for more advanced support in existing virtual whiteboards.

Finally, better support for the transition from sketches to other, downstream artifacts was also identified as an important feature to be developed. 
%"A digital whiteboard that would capture gestures or touches to facilitate that meeting and turn the artifact into digital, but maybe going further, "
\textit{"If we had a smarter way of recording it on whiteboard, as if it was a whiteboard, using simple scratches and boxes, this would automatically be transformed or suggested a certain transformation to the digital artifact as if it had constructed classes or class diagrams or component diagrams or sequence diagrams based on their whiteboard scratches. Perhaps, what is still missing is to understand that this transition is necessary, because the whiteboard is extremely useful, but often badly recorded, we cannot search it, we cannot verify what was decided at the previous meeting for the which is being decided now, we cannot compare the decisions and know that you've been wrong for three months and that three-month error could be recorded somewhere. So I think the transition for the digital should be smarter, something like: "look, you are trying to write a structure that some time ago you have already defined, you do not want to reuse what you did? Or you're trying to make an implementation much like another system that did the same thing in a particular architecture sharing repository, let's say."} We make two observations about this desire and particular comment. First, we note that it once more points to the prior point of needing better facilities to search for past content. Second, we note that part of what is being asked for already exists: current virtual whiteboards can turn hand-drawn sketches into more formal diagrams and representations that can then be exported to various other tools. The comment, however, seeks a much deeper integration, one in which such more formal documents and even prior sketches are fed back into the virtual whiteboard experience to more deeply assist the architects at work. Other comments similarly highlighted the need for not just supporting designing in the right notation, but to actually offer more `smart support' for the activities at hand.

\textbf{Education.} The insights we garnered provide fertile ground for how students are educated in the topic of software architecture as well. Beyond the need to cover architecture as a separate topic (which many programs do only peripherally so, although exceptions exist~\cite{4698906, 10.1145/3017680.3017737}), perhaps the most important factor is to teach the importance of architecture meetings at the whiteboard: what is typically discussed, how to conduct them, what kinds of perspectives should be brought to the discussion, the role of sketches in supporting the discussion taking place, and how to take those sketches into further development activities, and more. Given that the architecture forms the core of any system, and given that much of architecture design and refinement takes place at the whiteboard, the need for careful consideration how to teach these topics best is high. Existing courses on how to design software (e.g.,~\cite{DBLP:conf/icse/BenedettoN20, DBLP:conf/icse/Li20, DBLP:conf/icse/RukmonoC22}) as well as software maintenance (e.g.,~\cite{4385252, 8919211}) might provide both inspiration and serve as potential starting points in this regard. 

%we discussed along this study can be incorporated in software architecture courses. Initially, educators can exducaplain the importance of the meetings, topics commonly discussed in the meetings, types of sketches created, design decisions taken, decisions preserved/not preserved from sketch to code, aspects that change from sketch to implementation and their reasons. Purposes, approaches and problems with documentation should also be taught.

Beyond traditional courses in degree programs, we also suggest more advanced, turn-key architecture courses in which the human and social aspects of meetings are explored side-by-side with the technical considerations that go into architecture design and evolution.  Topics in such a course should include  team composition, the importance of mixed level participation, proper meeting organization and conduct, and perhaps particularly what architects can do to ensure that less information is missing from the meetings they conduct.
%as well as what represents a good and productive meeting; and what software architects consider missing from the meetings. 

\section{Threats to Validity}

In this section, we discuss several threats to validity for our study.

\textbf{Conclusion Validity.} Threats to conclusion validity are concerned with issues that relate to the treatment and the outcomes of the study, including, for instance, the choice of sample size and, as another example, the care taken in the implementation of a study \cite{10.5555/2349018}. In our work, we conducted interviews with open-ended questions in which the participants were asked to provide their perceptions and point-of-views. The interviews were then corroborated through a survey. The interviews were conducted at 18 different companies and when they happened within the same company, the participants were warned not talk to each other about it to avoid bias. In addition, we requested and were given access to experienced software architects at each company, to avoid the interviewees not possessing the necessary deep and long-term experience and knowledge in our area of investigation. We approached the design and implementation of the survey with the same level of care. Another aspect that is critical for conclusion validity is the quality of the material used in the study. Thus, to ensure that the interview prompt and survey instrument were of high quality, a pilot interview was conducted with a software architect and a survey pre-testing was performed with two software architects. Finally, to avoid the threat of drawing false conclusions based on the interview data, we carefully validated our interviews and findings with the participants as we performed analysis, asking for clarification when so needed. 

\textbf{Internal Validity.} 
%Our interview and analysis processes contain threats to internal validity. 
To reduce introducing interviewer bias during the interviews, we kept our questions open-ended and let participants talk most of the time. Additionally, it is possible that the participants might not have mentioned some points that, given more time to think, they could have brought up. To ameliorate this, we concluded the interviews by asking the participants whether they had any further thoughts and gave them ample time to respond before concluding the interviews. Similarly, we concluded the survey with a question as to whether survey participants had any additional thoughts they wanted to share. Interviewing participants remotely might also introduce some bias as compared to interviewing in person, for instance, by the interviewees giving shorter, incomplete, or unclear answers. We attempted to reduce this bias by following up with the interviewees if we felt an answer needed more clarification for us to be able to understand it in retrospect. 
%In most of the cases, participants thoroughly answered our questions. 
Because they were derived from the answers from the interviews, a possibility exists that the questions on the survey might not have been sufficiently representative (e.g., additional aspects that change from sketch to implementation, additional approaches to document whiteboard software architecture meetings). This was mitigated by the ability for the survey participants to provide additional thoughts through an open-ended question at the conclusion of the survey.
Finally, while our analysis was systematic, other researchers may discern different aspects than ours.

\textbf{Construct Validity.} There are threats to construct validity from the lack of a clear definition of a software architect. Nevertheless, in general, participants understood that we meant an experienced member in the development team responsible for making high-level design choices, validating them, and communicating those decisions to relevant stakeholders. In addition, we verbally clarified whenever there appeared to be some kind of confusion, both at the start of the interviews and throughout. Another threat to construct validity is related to the potential problem of evaluation apprehension \cite{10.5555/2349018}. It was mitigated by letting the participants know they would remain anonymous as well as by assuring them that all information gathered during the interviews and survey would solely be used only by the research team and never shared beyond.

\textbf{External Validity.} Our 27 interviews were conducted with software architects working in 18 different companies. Though these interviews yielded important insights, it can be considered a small sample. In addition, we only sampled software architects from five countries and findings may not generalize to other countries and companies. The same threat exists concerning the participants in the survey. Even though the respondents reside in 9 countries across four continents, our findings may not generalize to represent the experiences and perceptions of all software architects.

\section{Conclusion}
%Growing acknowledgement that some activities are better done in person.  Architecture design is one of them.

Becoming a software architect takes time and effort. In addition to having serious technical responsibilities in being the person who is primarily responsible for the software architecture and making architectural decisions, a software architect is also responsible for the many social aspects involved in the design and implementation of the architecture, including the subject of our study: conducting whiteboard software architecture meetings and bridging the outflow from these meetings to implementation. To date, such meetings have not been studied in detail and the realities of transferring results from the meetings to implementation also are not fully understood yet.
%guiding the development team through the implementation of the architecture and dealing with non-technical aspects (negotiation, facilitation, politics, and so on). 

In this paper, we contribute a mixed qualitative and quantitative study to investigate software architects' perceptions on whiteboard software architecture meetings. Based on interviews with 27 experienced architects and a subsequent survey with an additional 46 experienced software architects, our study yields twelve observations that range from reasons why software architects go to the whiteboard and perspectives on including experts and novices in the meetings, through how they document the outcomes of the meetings and why they document, to the kinds of changes they witness when the outcomes of the whiteboard meetings transition to implementation and the reasons for those changes. Our study is the first study of this kind, with the findings giving rise to further study, offering concrete advice for practicing architects, providing guidance for future tool design, and suggesting new topics for educating future software architects.

%provided a list of 12 observations including technical, human, and social perspectives of the meetings. These observations are being used as input in our research group to conduct empirical studies and develop new tools.

\section{Acknowledgments}
We thank all the software architects who participated in our interviews and survey.

\bibliographystyle{IEEEtran}

\begin{IEEEbiography}{Michael Shell}
Biography text here.
\end{IEEEbiography}

\begin{IEEEbiographynophoto}{John Doe}
Biography text here.
\end{IEEEbiographynophoto}

%\newpage

\begin{IEEEbiographynophoto}{Jane Doe}
Biography text here.
\end{IEEEbiographynophoto}

\end{document}